\documentclass[twocolumn,aps,prd,preprintnumbers,superscriptaddress,nofootinbib,amsmath,amssymb,floats,floatfix,notitlepage,longbibliography]{revtex4-1}

\usepackage{orcidlink}
\usepackage{lipsum}
\usepackage{graphicx}
\usepackage{subfigure}
\usepackage{braket}
\usepackage[utf8]{inputenc}
\usepackage{sans}
\usepackage{changes}
\usepackage{hyperref}
\hypersetup{colorlinks=true,linkcolor=blue,urlcolor=blue,citecolor=blue}
\usepackage[toc,page]{appendix}
\usepackage[normalem]{ulem}
\usepackage{adjustbox}
\usepackage{latexsym}
\usepackage{amsmath}
\usepackage{amssymb}
\usepackage{amsfonts}
\usepackage{dcolumn}
\usepackage{bm}
\usepackage{tikz}
\usepackage{bigints}
\usepackage{array,tabularx,multirow,booktabs}
\usepackage[tracking=true]{microtype}
\SetTracking{}{500}
\SetTracking{encoding={*}, shape=sc}{40}
\UseRawInputEncoding 
\allowdisplaybreaks

\newcommand{\dd}{\mathrm{d}}

\newcommand{\abs}[1]{\left|#1\right|}

\begin{document} \sloppy

\title{Topological Signatures and Geometrothermodynamics of Critical Phenomena in Regularized Maxwell Black Holes}


\author{Y. Sekhmani\orcidlink{0000-0001-7448-4579}}
\email[Email: ]{sekhmaniyassine@gmail.com}
\affiliation{Center for Theoretical Physics, Khazar University, 41 Mehseti Street, Baku, AZ1096, Azerbaijan}
\affiliation{Centre for Research Impact \& Outcome, Chitkara University Institute of Engineering and Technology, Chitkara University, Rajpura, 140401, Punjab, India}
\affiliation{Institute of Nuclear Physics, Ibragimova, 1, 050032 Almaty, Kazakhstan}
\affiliation{Department of Mathematical and Physical Sciences, College of Arts and Sciences, University of Nizwa, P.O. Box 33, Nizwa 616, Sultanate of Oman}
\author{G.~G.~Luciano\orcidlink{0000-0002-5129-848X}}
\email[Email: ]{giuseppegaetano.luciano@udl.cat }
\affiliation{Departamento de Qu\'{\i}mica, F\'{\i}sica y Ciencias Ambientales y del Suelo, Escuela Polit\'ecnica Superior -- Lleida, Universidad de Lleida, Av. Jaume II, 69, 25001 Lleida, Spain
}
\author{S.K. Maurya\orcidlink{0000-0003-0261-7234}}  \email[Email: ]{sunil@unizwa.edu.om}
\affiliation{Department of Mathematical and Physical Sciences, College of Arts and Sciences, University of Nizwa, P.O. Box 33, Nizwa 616, Sultanate of Oman}
\author{J. Rayimbaev\orcidlink{0000-0001-9293-1838}
}
\email[Email: ]{javlon@astrin.uz}
\affiliation{Urgench State University, Kh. Alimjan Str. 14, Urgench 221100, Uzbekistan}
\affiliation{University of Tashkent for Applied Sciences, Str. Gavhar 1, Tashkent 100149, Uzbekistan}
\affiliation{National University of Uzbekistan, Tashkent 100174, Uzbekistan}
\affiliation{Tashkent State Technical University, Tashkent 100095, Uzbekistan}

\author{M. K. Jasim}
\email[Email: ]{mahmoodkhalid@unizwa.edu.om}\affiliation{Department of Mathematical and Physical Sciences, College of Arts and Sciences, University of Nizwa, Nizwa 616, Sultanate
of Oman}
\author{I. Ibragimov} 
\email[Email: ]{i.ibragimov@kiut.uz}
\affiliation{Kimyo International University in Tashkent, Shota Rustaveli street 156, Tashkent 100121, Uzbekistan}
\author{S. Muminov}
\email[Email: ]{sokhibjan.muminov@mamunedu.uz}
\affiliation{Mamun University, Bolkhovuz Street 2, Khiva 220900, Uzbekistan}

\begin{abstract}
We study the thermodynamic topology and microscopic interaction properties of charged black holes in RegMax gravity, focusing on the role of the coupling parameter $\alpha$. Using the Duan topological current method together with Ruppeiner geometry, we show that $\alpha$ controls a sharp change in phase structure. Above a certain critical threshold, we find that the Duan defect curve develops an intermediate branch and vertical tangency points, producing continuous (second-order) critical behaviour. Furthermore, the Ruppeiner curvature becomes negative at very small horizon radii before turning positive and progressively vanishing at larger radii. By contrast, below the critical value of the coupling,
the intermediate black hole phase disappears, and the system shows a simpler small/large first-order/coexistence behaviour driven by free-energy competition. In this regime, the Ruppeiner curvature remains predominantly positive. Overall, increasing $\alpha$ enriches the thermodynamic topology (allowing for second-order criticality) while simultaneously reducing the domain in which classical energy conditions (ECs) are satisfied, thus linking exotic thermodynamic behaviour to more severe violations of standard energy conditions.
\end{abstract}


\maketitle


\section{introduction}
The study of black hole thermodynamics provides a central subject at the crossroads of gravitation, quantum mechanics and statistical physics. Foundational contributions by Bekenstein and Hawking revealed that black holes carry an entropy proportional to the area of their event horizon and radiate thermally with a temperature set by their surface gravity~\cite{Bekenstein:1973ur,Hawking:1975vcx}. Within asymptotically AdS geometries, these thermodynamic variables display striking similarities to those of ordinary matter systems, including phase structures analogous to the Van der Waals fluid in the case of Reissner--Nordstr\"om AdS black holes~\cite{Kubiznak:2012wp}. Related developments extend this analogy to broader non-extensive statistical settings~\cite{Rani:2023jai,Rani:2022xza,Luciano:2023fyr,2025PDU....4801876T,Luciano:2023bai,Nakarachinda:2022gsb} and deformed uncertainty relation frameworks~\cite{Gangopadhyay:2013ofa,Hassanabadi:2021kuc,Kim:2007hf,Sekhmani:2025zwc}.  For further studies on black hole thermodynamics, see e.g.~\cite{Chamblin:1999hg,Caldarelli:1999xj,Padmanabhan:2009vy,Cai:2005ra,Wald:1999vt,Kubiznak:2016qmn,Dolan:2011xt,Cvetic:2010jb,Gunasekaran:2012dq,Hayward:1997jp,Capozziello:2022ygp,Ghaffari:2025qmv}.

To investigate black hole thermodynamics and phase transitions from a topological standpoint, a largely adopted method is Duan's $\phi$-mapping framework \cite{Duan:1998it}, which reformulates the properties of the thermodynamic scalar potential in terms of a vector field defined on the \((T,r_h)\) plane, where $T$ and $r_h$ denote the black hole temperature and horizon radius, respectively. The zeros of this vector field signify thermodynamic critical points and are characterized by integer winding numbers: \(+1\) for standard phase transitions and \(-1\) for novel or
inverse ones. This methodology establishes a stable topological classification that enriches and extends conventional thermodynamic descriptions.

In contrast to standard methods based on the study of response functions (e.g., heat capacity singularities) or on the analysis of thermodynamic potentials, Duan's topological approach presents several significant advantages. Most notably, it furnishes a coordinate-invariant and metric-independent classification of critical points via topological invariants, ensuring robustness under reparametrizations or deformations of the thermodynamic phase space. In addition, by associating integer winding numbers with critical points, this framework encodes not only their position but also their qualitative character, thereby distinguishing between ordinary and inverse phase transitions in a mathematically precise way. A further strength is that the construction remains applicable even when conventional thermodynamic observables lose analyticity or fail to show clear divergences, thus broadening the scope of phase structure investigations to regimes that are inaccessible with traditional techniques~\cite{Wei:2022dzw,Wei:2022dzw2}. {For further applications of thermodynamic topology in relation to various types of black holes, the reader is directed to the Refs. \cite{NooriGashti:2024gnc,Anand:2025ttx,Anand:2025mlc,Afshar:2025oqn,Anand:2025qow}}

Complementary to these topological methods, a geometric perspective on thermodynamics is provided by the Weinhold and Ruppeiner formalisms~\cite{Wein1,Rupp1,Rupp2}. In this framework, one assigns a Riemannian structure to the thermodynamic phase space, where the scalar curvature of the corresponding metric encodes information about the underlying microscopic interactions. In particular, the sign of the Ruppeiner curvature is often interpreted as indicating whether the dominant interactions are repulsive, while its divergences typically signal the presence of phase transitions. This geometric approach thus offers an intuitive microscopic interpretation of black hole thermodynamics and can be fruitfully combined with topological analyses to achieve a more complete characterization of critical phenomena. {For more applications of thermodynamic geometry to different kinds of black holes, the reader is referred to Refs. \cite{Wei:2012ui,Zhang:2014uoa,Hendi:2014kha,Hendi:2015xya, Zhang:2015ova,EslamPanah:2018ums,Ghosh:2019pwy,HosseiniMansoori:2020yfj}.}

While these advances provide powerful tools to classify black hole phase transitions from a thermodynamic and topological perspective, the underlying matter content sourcing such geometries plays an equally crucial role. In particular, nonlinear electrodynamics (NLE) has emerged as a natural framework for addressing long-standing issues, such as the removal of divergences in point-charge fields and the construction of regular black holes. Among the various NLE models, the Born--Infeld theory~\cite{Born:1934gh} stands as the historical prototype, later generalized to string-theoretic and D-brane contexts~\cite{Fradkin:1985qd,Leigh:1989jq}, and more recently employed in regular black hole physics~\cite{Ayon-Beato:1998hmi}.  

Within this broader landscape, a particularly interesting model is the so-called \emph{Regularized Maxwell} (RegMax) theory (see Ref.~\cite{Hale:2023dpf} and references therein). This framework smoothly reduces to Maxwell electrodynamics in the weak-field regime while implementing a minimal regularization of the point-charge self-energy. Beyond these appealing features, RegMax stands out for its gravitational implications: it is the only NLE model depending solely on the invariant \(F_{\mu\nu}F^{\mu\nu}\) that admits radiative solutions in the Robinson--Trautman class~\cite{Tahamtan:2020lvq}, generalizing earlier non-radiative cases~\cite{Tahamtan:2015bha}. Remarkably, unlike their Maxwell counterparts, these solutions are well-posed. In addition, RegMax accommodates slowly rotating black holes in close analogy with the Maxwell case~\cite{Kubiznak:2022vft}, and even supports natural generalizations of the C-metric, thus providing charged and accelerated black holes within its framework.

Building on these lines of research, in this work, we explore the thermodynamic topology and microscopic interaction properties of charged AdS black holes in RegMax gravity. Specifically, we focus on the role of the coupling parameter $\alpha$, which characterizes the RegMax Lagrangian and quantifies the deviation from standard Maxwell electrodynamics. Depending on the value of this constant, we show that distinct regimes emerge, each endowed with different thermodynamic and microstructural properties. This makes the exploration of the $\alpha$-parameter space not only a probe of nonlinear electromagnetic effects, but also a window into unprecedented patterns of black hole criticality.

The structure of the paper is as follows. In the next section we analyze the charged black hole solutions of RegMax theory. Section~\ref{topology} is devoted to the study of thermodynamic topology, while Section~\ref{Sectthermalgeometry} addresses the geometrothermodynamic aspects. We conclude with a summary of results and outlook in Sec.~\ref{sec:theend}. Throughout this work, we adopt natural units in which $\hbar = c = G = k_{B} = 1$.

\section{Charged black hole solution}
\label{sec:bhsolution}
In this section, we focus on geometric curvature singularities, the three-dimensional embedding derived from the lapse function solution $f(r)$, and the analysis of ECs in the context of black holes coupled to the RegMax term within the RG framework. We then investigate the thermodynamic topology and the geothermodynamic processes (i.e., heat transfer) governed by the Ruppeiner-Ricci scalar, considering three regimes: subcritical, critical, and supercritical bounds.

\subsection{Spacetime solution}
The charged black hole solution within the RegMax theory has been developed as detailed in \cite{Hale:2023dpf}. It is expressed in the following standard form: 
\begin{equation}
\mathrm{d} s^2=g_{\mu\nu}\mathrm{d} x^\mu \mathrm{d} x^\nu=-f(r)\mathrm{d} t^2 +\frac{\mathrm{d} r^2}{f(r)}+r^{2}\left(\mathrm{d} \theta^2+\sin^2\!\theta \mathrm{d} \varphi^2\right),  
\label{metric}
\end{equation}
where the lapse function $f(r)$ is explicitly given by
\begin{eqnarray}
f(r)&=&1-2\alpha^{2}|Q|+\frac{4\alpha |Q|^{3/2}-6M}{3r}+\frac{r^2}{\ell^2} \\
\nonumber
&&+4\alpha^{3}r\sqrt{|Q|}-4r^2\alpha^4\ln\left(1+\frac{\sqrt{|Q|}}{r\alpha}\right)\label{eq:f}\\
\nonumber
&=&{1-\frac{2M}{r}+\frac{Q^2}{r^2}+\frac{r^2}{\ell^2}+\frac{Q^2}{r^2}\sum_{n=1}^{\infty}\frac{4}{n+4}\Bigl(\frac{-\sqrt{|Q|}}{\alpha r}\Bigr)^n},
\end{eqnarray}
where $Q$ is the (asymptotic) electric charge, $M$ is the black hole mass, $\alpha$ is the RegMax coupling parameter that governs the deviation from standard Maxwell electrodynamics\footnote{{{Working in units $\hbar=c=G=k_B=1$, the threshold $\alpha^2|Q|$ is dimensionless; therefore $[\alpha]=[|Q|]^{-1/2}$}}.}, and $\ell$ is the AdS radius related to the cosmological constant by $\Lambda=-3/\ell^{2}$ (see below for technical details).

Since the matter sector supports a charged source within the RegMax theory, the corresponding vector potential associated with the metric takes the form
\begin{eqnarray}
A=-\frac{Q\alpha}{r\alpha+\sqrt{\abs{Q}}} \dd t\,, \label{eq:potential}
\end{eqnarray}
resulting in an appropriate field strength in the form of
\begin{eqnarray}
F= \mathcal{E} dr \wedge \dd t\,,\quad
\mathcal{E} \equiv \frac{Q\alpha^2}{(r\alpha+\sqrt{\abs{Q}})^2}\,,  \label{eq:eli}
\end{eqnarray}
and is characterized by the following two invariants:
\begin{eqnarray}
\mathcal{S}=\frac{Q^2\alpha^4}{(r\alpha+\sqrt{\abs{Q}})^4}
\,,\quad \mathcal{P}=0\,.    
\end{eqnarray}

The provided solution is static, spherically symmetric, and exhibits a singularity at $r=0$. Within a specific range of parameters $\{M,Q,\alpha, \ell\}$, it characterises a charged black hole. The horizon radius $r_h$ is determined as the largest root of the equation $f(r=r_h) = 0$. For small masses $M$, the behaviour of $f(r)$ closely mimics that of the Reissner--Nordstr{\"o}m solution. Clearly, as the mass increases from zero, the scenario evolves from having no roots, which corresponds to a naked singularity, to a single degenerate root indicative of an extremal black hole. Ultimately, this progression leads to the emergence of two roots, characteristic of a non-extremal black hole, which feature inner and outer horizons. In the case of masses that are larger, i.e. 
\begin{widetext}
    \begin{figure}[htbp]
  \centering
\includegraphics[width=\textwidth]{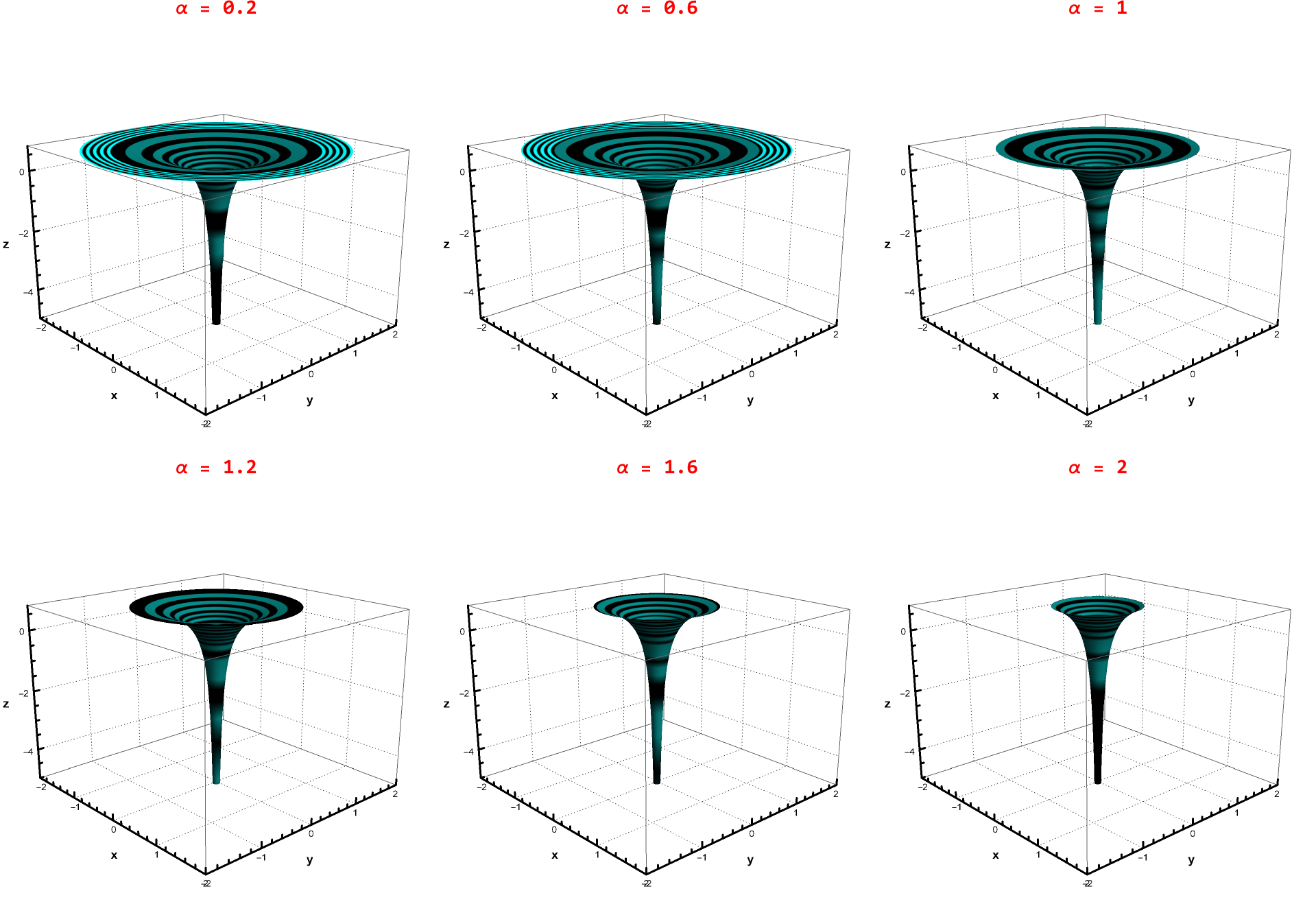}
  \caption{Embedding-style surfaces \(z=f(r)\) for \(Q=0.1\), \(M=0.2\), \(\Lambda=-0.001\).
  Each panel corresponds to one value of \(\alpha\in\{0.2,0.6,1.0,1.2,1.6,2.0\}\) (panels arranged left-to-right, top-to-bottom). 
  }
\label{fig:embedding_panels}
\end{figure}
\end{widetext}
\begin{equation}
M>M_{\mbox{\tiny marginal}}=\frac{2\alpha\abs{Q}^{3/2}}{3}\,,
\end{equation}
the behaviour of the metric function transitions from the Reissner--Nordstr{\"o}m regime to a Schwarzschild-like one, characterised by the presence of a single non-extremal horizon.

\subsection{Embedding analysis}
To grasp this compact gravitational object better, it is worth studying the three-dimensional embedding based on the lapse function $f(r)$. This is done with fixed parameters set at $Q=0.1$ with $M=0.2$ and $\Lambda=-0.001$. Thus, Fig.~\ref{fig:embedding_panels} illustrates the surface $z=f(r)$ represented on the domain $r (x,y,z)\in[0.01,2.0]$  for the six values $\alpha=\{0.2,0.6,1.0,1.2,1.6,2.0\}$. In numerical terms, Table~\ref{tab:root-summary} presents the numerically determined outer horizon radii $r_h$ (the root of $f(r)=0$ in the displayed range) as well as illustrative values of $f(r)$ at the boundaries of the displayed interval.
\begin{table}[htbp]
  \centering
  \caption{Numeric summary for the slice $Q=0.1$, $M=0.2$, $\Lambda=-0.001$. Values were computed on the same $r$-grid used to produce the panels.}
  \label{tab:root-summary}
  \begin{tabular}{c c c c}
    \hline\hline
    $\alpha$ & horizon $r_h$ (root $f(r_h)=0$) & $f(r_{\min}=0.01)$ & $f(r_{\max}=2.0)$ \\
    \hline
    0.2 & 0.392183 & -38.160628 & 0.806875 \\
    0.6 & 0.373782 & -36.503652 & 0.839401 \\
    1.0 & 0.348381 & -34.872375 & 0.903554 \\
    1.2 & 0.333468 & -34.065242 & 0.947596 \\
    1.6 & 0.301030 & -32.465948 & 1.059651 \\
    2.0 & 0.267495 & -30.884128 & 1.203685 \\
    \hline\hline
  \end{tabular}
\end{table}

A closer inspection reveals that for each choice of $\alpha$ indicated, the function $f(r)$ is monotonically increasing on $r\in[0.01,2]$ and crosses zero precisely once. In this domain, the spacetime presents a single (outer) horizon at $r_h$, as is shown in Table~\ref{tab:root-summary}. Furthermore, increasing $\alpha$ (at fixed $Q, M, \Lambda$) brings the zero of $f(r)$ closer to smaller radii: the outer horizon decreases from $r_h\approx0.392$ at $\alpha=0.2$ to $r_h\approx0.267$ at $\alpha=2.0$.

On the other hand, the inverse--radius term, proportional to $(4\alpha|Q|^{3/2}-6M)/(3r)$, dominates in the limit $r\to 0$. For the chosen model ($Q=0.1$, $M=0.2$), this coefficient is negative for all values of $\alpha$ considered, producing a strongly negative behavior near $r\approx 0$. At large $r$, the positive linear contribution $4\alpha^{3}\sqrt{|Q|}\,r$ competes with the negative, effectively quadratic term $-4\alpha^{4}r^{2}\ln\!\left(1+\tfrac{\sqrt{|Q|}}{\alpha r}\right)$. Within the range shown, the overall effect is an increase in $f(r)$, leading to positive values at $r=2$. Finally, the boundary height in each panel grows with $\alpha$, while the depth of the central well becomes shallower as $\alpha$ increases. Moreover, the intersection of the surface with the $z=0$ plane (the horizon) shifts toward smaller $r$.

\subsection{Kretschmann scalar}
At this stage, we employ curvature singularity diagnostics to investigate the physical properties encoded in the spacetime solution. In this context, the Kretschmann scalar serves as a powerful tool, as the divergence of $\mathcal{K}=R^{\alpha\beta\gamma\delta}R_{\alpha\beta\gamma\delta}$ signals the presence of a scalar curvature singularity. The explicit form of the Kretschmann scalar corresponding to the spacetime solution \eqref{eq:f} is therefore given by
\begin{widetext}
    \begin{eqnarray}
R^{\alpha\beta\gamma\delta}R_{\alpha\beta\gamma\delta}  &=&\left[\frac{8 \alpha  | Q| ^{3/2}}{3 r^3}+\frac{4 \alpha ^4 | Q| }{\left(\sqrt{| Q| }+\alpha  r\right)^2}+\frac{8 \alpha ^4 \sqrt{| Q| }}{\sqrt{| Q| }+\alpha  r}-8 \alpha ^4
   \log \left(\frac{\sqrt{| Q| }}{\alpha  r}+1\right)+\frac{2}{\ell^2}-\frac{4 M}{r^3}\right]^2\nonumber\\
   &+&\frac{4 \left[-\frac{4 \alpha  | Q| ^{3/2}}{3 r^2}+4 \alpha ^3 \sqrt{| Q| }
   \left(\frac{\alpha  r}{\sqrt{| Q| }+\alpha  r}+1\right)-8 \alpha ^4 r \log \left(\frac{\sqrt{| Q| }}{\alpha  r}+1\right)+\frac{2 r}{\ell^2}+\frac{2
   M}{r^2}\right]^2}{r^2}\nonumber\\
   &+&\frac{4 \left[\frac{4 \alpha  | Q| ^{3/2}-6 M}{3 r}-\alpha ^2 | Q| -4 \alpha ^4 r^2 \log \left(\frac{\sqrt{| Q| }}{\alpha  r}+1\right)+4 \alpha ^3 r
   \sqrt{| Q| }+\frac{r^2}{\ell^2}\right]^2}{r^4}\,.
    \end{eqnarray}
\end{widetext}
The Kretschmann invariant displays the anticipated central curvature singularity and an asymptotic behaviour resembling AdS space at infinity. A short-distance expansion reveals a principal power-law divergence, as the expression below shows
\begin{equation}
R^{\alpha\beta\gamma\delta}R_{\alpha\beta\gamma\delta}\;\xrightarrow[r\to0]{}\;\frac{16}{3}\,\frac{\big(2\alpha\,|Q|^{3/2}-3M\big)^{2}}{r^{6}}+O(r^{-4}),
\end{equation}
so generically the invariant blows up as $r^{-6}$. The coefficient is an exact square, and thus the leading $r^{-6}$ term vanishes only under the non-generic fine-tuning condition $3M=2\alpha|Q|^{3/2}$. In this case, the subleading $r^{-4}$ contributions together with logarithmic terms (involving factors such as $\log r$) persist and must be examined to assess regularity. At large $r$, the invariant asymptotically approaches the AdS value in such a way that
\begin{equation}
R^{\alpha\beta\gamma\delta}R_{\alpha\beta\gamma\delta}\;\xrightarrow[r\to\infty]{}\; -\frac{8}{\ell^2}\;+\;O(1/r),
\end{equation}
so the spacetime is asymptotically AdS.  Practically speaking, a higher value of $\alpha$ or $|Q|$ magnifies the curvature near the centre (through the factors $|Q|^{3/2}$ and $\alpha$ in the leading coefficient), and the logarithmic elements $\alpha^{4}\log(1+\sqrt{|Q|}/(\alpha r))$ introduce a smoother but noticeable structure at intermediate radii. Note that Kretschmann's scalar remains finite for any non-zero horizon radius $r_h>0$ and diverges only at the central point $r=0$.

\subsection{Physical properties}
As demonstrated in \cite{Hale:2023dpf}, the black hole solution described above is characterized by the following physical parameters: the asymptotic mass $M$ and the electric charge $Q$, which are given by
\begin{eqnarray}
\label{M}
    M&=&\frac{1}{6} \Bigg\{3 r_h \left[-4 \alpha ^4 r_h^2 \log \left(\frac{\sqrt{| Q| }}{\alpha  r_h}+1\right)+\frac{r_h^2}{l^2}+1\right]\nonumber\\
    &+&4 \alpha  | Q| ^{3/2}+12 \alpha ^3 r_h^2 \sqrt{| Q| }-3
   \alpha ^2 r_h | Q| \Bigg\}\,,\\ Q&=&\frac{1}{4\pi}\int_{S^2}*D\,,   \label{Q RegMax}
\end{eqnarray}
Similarly, for the black hole temperature $T$ and entropy $S$, we have
\begin{eqnarray}
    T&=&\frac{f'(r_h)}{4\pi}\nonumber\\
&=&\frac{\alpha r_h(6|Q|\alpha^2+1)-2|Q|^{3/2}\alpha^2+\sqrt{|Q|}(1+12\alpha^4r_h^2)}{4\pi r_h(\alpha r_h+\sqrt{|Q|})}\nonumber\\
&&-\frac{3r_h\alpha^4}{\pi}\log\bigl(1+\frac{\sqrt{|Q|}}{r_h\alpha}\bigr)+\frac{3r_h}{4\pi \ell^2}\,,\label{Tform}\\
S&=&\frac{\mbox{Area}}{4}=\pi r_h^2\,,\label{SS}
\end{eqnarray}
and the electrostatic potential reads
\begin{equation}
    \label{phiTD}
\phi=-\xi\cdot A\Bigr|_{r=r_h}=\frac{\alpha Q}{\alpha r_h+\sqrt{|Q|}}\,, 
\end{equation}
where, in the last formula, we have used the fact that the horizon is generated by the Killing vector field $\xi=\partial_t$.

Finally, since the solution is asymptotically AdS, we can consider the corresponding pressure-volume term \cite{Kastor:2009wy, Kubiznak:2016qmn}, 
\begin{equation}
    \label{Vform}
P=-\frac{\Lambda}{8\pi}=\frac{3}{8\pi \ell^2}\,,\quad  V=\Bigl(\frac{\partial M}{\partial P}\Bigr)_{S,Q,
\alpha}=\frac{4}{3}\pi r_h^3\,,
\end{equation}
and the ``{\em $\alpha$-polarization potential}'' \cite{Gunasekaran:2012dq}
\begin{eqnarray}
   \mu_\alpha&=&\Bigl(\frac{\partial M}{\partial \alpha}\Bigr)_{S,Q,P}\nonumber\\
&=&-\frac{2}{3}\frac{2|Q|^{3/2}\alpha r_h-Q^2-12\alpha^3r_h^3\sqrt{|Q|}-6|Q|\alpha^2r_h^2}{r_h\alpha+\sqrt{|Q|}}\nonumber\\
&&-8\alpha^3 r_h^3\log\Bigl(1+\frac{\sqrt{|Q|}}{r_h\alpha}\Bigr)\,,  
\end{eqnarray}
reflecting the fact that $\alpha$ is a dimension-full quantity.  

It is then easy to verify that the above quantities obey the extended first law and the corresponding Smarr relation  \cite{Hale:2023dpf}:
\begin{eqnarray}
    \delta M&=&T\delta S+\phi \delta Q+V\delta P+\mu_\alpha \delta \alpha\,,\\[2mm]
M&=&2TS+\phi Q-2VP-\frac{1}{2}\mu_\alpha \alpha\,.
\label{M2}
\end{eqnarray}
Moreover, the corresponding canonical (fixed charge) and grandcanonical (fixed potential) ensembles feature various critical points and phase transitions, see 
\cite{Hale:2023dpf}. 

In what follows, we examine the energy conditions for the black hole solution.

\subsection{Energy Conditions}
The elements of the stress-energy tensor $T_{\mu\nu}$ governed by Einstein equations for RegMAx BHs with a negative cosmological are as follows:
        \begin{align}
        \rho&=\frac{\alpha ^2 \sqrt{| Q| } \bigg[| Q| -\alpha  r \bigg(7 \sqrt{| Q| }+12 \alpha  r\bigg)\bigg]}{r^2 \left(\sqrt{| Q| }+\alpha  r\right)}\nonumber\\
        &+12 \alpha ^4 \log
   \left(\frac{\sqrt{| Q| }}{\alpha  r}+1\right) =-P_r\label{rho}\,,\\
      P_\theta&=P_\phi=2 \alpha ^3 \Bigg[\frac{\sqrt{| Q| } \left(9 \alpha  r \sqrt{| Q| }+2 | Q| +6 \alpha ^2 r^2\right)}{r \left(\sqrt{| Q| }+\alpha  r\right)^2}\nonumber\\
      &-6 \alpha  \log
   \left(\frac{\sqrt{| Q| }}{\alpha  r}+1\right)\Bigg].\label{P}
    \end{align}
    
        \begin{figure}[t!]
\centering
\includegraphics[width=7.5cm]{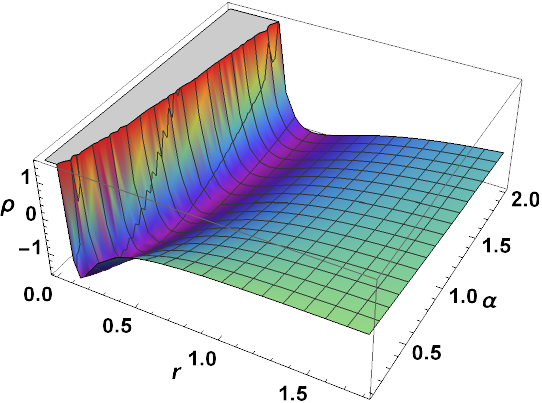}\\
\includegraphics[width=7.8cm]{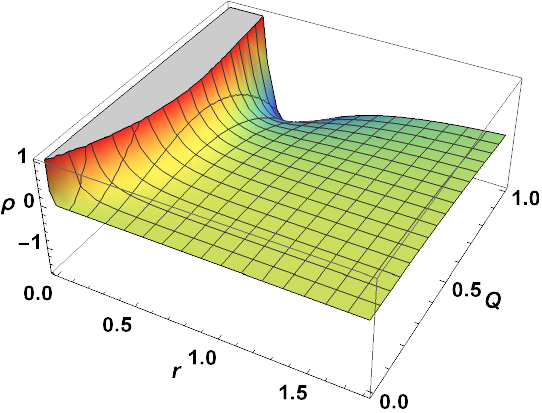}
\caption{Plot of WEC using the fixed benchmark $\alpha=1$ and $Q=1$.}
\label{wec}
\end{figure}

\begin{figure}[t!]
\centering
\includegraphics[width=7.5cm]{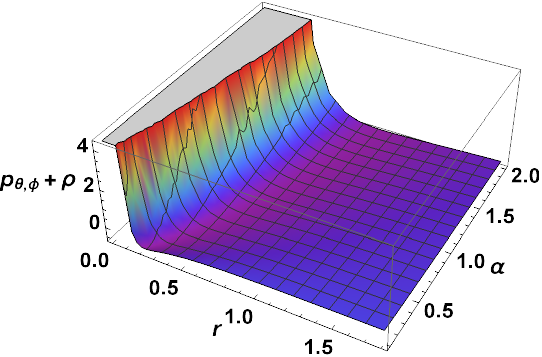}\\
\includegraphics[width=7.8cm]{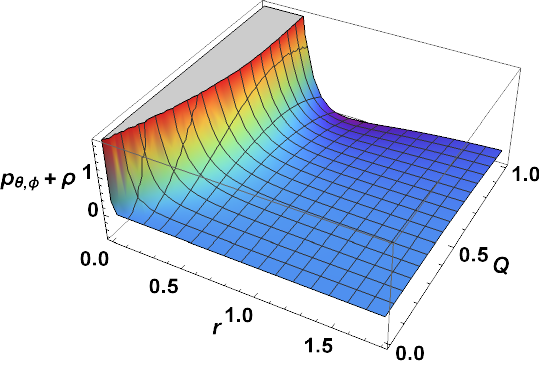}
\caption{Plot of NEC using the fixed benchmark $\alpha=1$ and $Q=1$.}
\label{nec}
\end{figure}

\begin{figure}[t!]
\centering
\includegraphics[width=7.5cm]{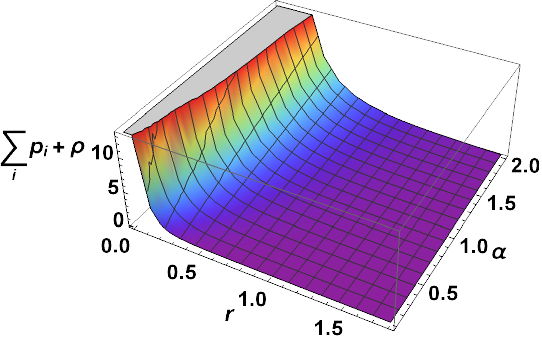}\\
\includegraphics[width=7.8cm]{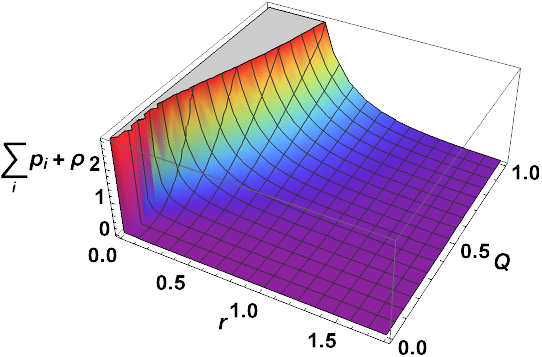}
\caption{Plot of SEC using the fixed benchmark $\alpha=1$ and $Q=1$.}
\label{sec}
\end{figure}

    \begin{itemize}
        \item The weak energy condition (WEC) requires that $T_{\mu\nu}\, t^\mu t^\nu\geqslant0$ everywhere, for any time vector $t^\mu$, which is equivalent to~\cite{Toshmatov:2017kmw}
    \begin{equation}
        \rho\ge0,\quad \rho+ P_i\ge0\quad (i=r, \theta, \phi)
    \end{equation}
    and so $\rho+P_r=0$ and
    \begin{align}
        &\rho+P_\theta=\rho+P_\phi\nonumber\\
        &=\frac{\alpha ^2 | Q|  \left[| Q| -\alpha  r \left(2 \sqrt{| Q| }+\alpha  r\right)\right]}{r^2 \left(\sqrt{| Q| }+\alpha  r\right)^2}
   \label{3-}
    \end{align}

For the spacetime solution, the WEC reduces to two nontrivial inequalities, 
\begin{equation}
\rho \geq 0, 
\qquad 
\rho + P_{\theta} \geq 0,
\end{equation}
since $P_{r} = -\rho$ makes $\rho + P_{r} \equiv 0$. Importantly, $\rho(r)$ diverges positively as $r \to 0^{+}$ 
(with leading behavior $\rho \sim \alpha^{2} |Q|/r^{2}$), so the condition $\rho \geq 0$ is always satisfied sufficiently close to the singularity. The angular combination admits a simple closedform expression,
\begin{equation}
\rho  \; \propto \; | Q| -\alpha  r \left(7 \sqrt{| Q| }+12 \alpha  r\right),
\end{equation}
which yields the exact critical radius
\begin{equation}
r_{\text{WEC}} =\frac{\left(\sqrt{97}-7\right) \sqrt{|Q|}}{24 \alpha }\,,
\end{equation}
such that
\begin{equation}
\rho  \geq 0 
\quad \Longleftrightarrow \quad 
0 < r \leq r_{\text{WEC}}.
\end{equation}
Thus the WEC holds only in the intersection of the small-$r$ region (see Fig. \ref{wec}) where $\rho \geq 0$ and the angular bound $r \leq r_{\text{WEC}}$. Practically, this means the WEC is satisfied in a neighbourhood of the origin but fails beyond $r_{\text{WEC}}$. Increasing $\alpha$ shrinks that neighbourhood; for example, with $|Q|=1$ one finds $r_{\text{WEC}} \approx 0.0593512$ for $\alpha=2$, but $r_{\text{WEC}} \approx 0.593512$ for $\alpha=0.2$.

\item The null energy condition (NEC) stipulates that $T_{\mu\nu}\, t^\mu t^\nu\geqslant0$ in the overall spacetime for any null vector $t^\mu$. The NEC predicts $\rho+P_r \geqslant 0$ which is identically zero, and $\rho+P_\theta=\rho+P_\phi \geqslant 0$. Analytically, the condition $T_{\mu\nu}\, t^\mu t^\nu\geqslant0$ reduces to the quadratic
\begin{equation}
-\,r\alpha\,(r\alpha+2\sqrt{|Q|})+|Q|=0,
\end{equation}
whose physically relevant root is
\begin{equation}
r_{\mathrm{NEC}}=\frac{\left(\sqrt{2}-1\right) \sqrt{| Q| }}{\alpha }.
\end{equation}
Thus NEC holds for $0<r<r_{\mathrm{NEC}}$ and is violated for $r>r_{\mathrm{NEC}}$. So
\begin{align}
\rho+p_{\theta,\phi}&\sim\alpha^2|Q|/r^2\to+\infty,\quad r\to0^{+} \\
    \rho+p_{\theta,\phi}&\sim-2\alpha\sqrt{|Q|}/r^3\to0^{-}\quad r\to\infty.
\end{align}

\begin{enumerate}
\item For $\alpha = 2$ the NEC is satisfied only for extremely small radii 
$r < 0.2071$ (see Fig. \ref{nec}). For all physically interesting horizon radii 
(the outer horizon typically $r_{h} \gtrsim \mathcal{O}(1)$ in our examples), 
the NEC is violated. This indicates that large $\alpha$ enhances violations 
of standard energy conditions at macroscopic scales. The effect stems from 
the stronger $\alpha$--dependent terms in the stress tensor 
($\alpha^{2}$, $\alpha^{3}$, $\alpha^{4}\log$), and explains why the geometry 
and phase structure depart more strongly from the Reissner-Nordström behaviour.  

\item For $\alpha = 0.2$ the NEC holds up to $r \approx 2.07$, so a wide domain 
of horizon radii satisfies both the NEC and WEC combinations (see Figs. \ref{wec}-\ref{nec}). In this regime the 
system is closer to a standard physically sensible matter profile. This again 
matches the simpler phase structure obtained for small $\alpha$.
\end{enumerate}

\item The strong energy condition (SEC) asserts that  $T_{\mu\nu}\, t^\mu t^\nu\geqslant 1/2 \,T_{\mu\nu} t^\nu t_\nu$ globally, for any time vector $t^\mu$ which assumes that~\cite{Toshmatov:2017kmw}
\begin{equation}
        \rho+\sum_i P_i=P_r+2\,P_\theta\ge0.
        \label{37}
    \end{equation}
Substituting the exact expressions yields a closed-form combination of rational terms and a logarithmic contribution
\begin{equation}
\log \!\left( 1 + \tfrac{|Q|}{\alpha r} \right).
\end{equation}
Two robust analytical facts follow: 
\begin{enumerate}
    \item As $r \to 0^{+}$ the $-\rho$ term dominates (since 
    $\rho \sim +\tfrac{\alpha^{2} |Q|}{r^{2}}$), so
    $-\rho + 2P_{\theta} \to -\infty$, and the SEC is always violated arbitrarily close to the singularity. 
    \item The logarithmic term enters $-\rho + 2P_{\theta}$ with a negative coefficient 
    $\propto -\alpha^{4}$, so larger $\alpha$ systematically makes the SEC harder to satisfy at intermediate or large radii (see Fig. \ref{sec}). 
\end{enumerate}
Therefore, SEC violations are generic near the core and tend to persist or expand for supercritical couplings.
    \end{itemize}
    
\noindent Broadly speaking,
a simple analytic estimate showed that the radial domain where the null (and weak) energy conditions hold is controlled by the coupling $\alpha$ and the charge scale $|Q|$.  Solving $\rho+P_\theta=0$ in the small-radius approximation yields the leading threshold
\begin{equation}
\label{rnec}
r_{\rm NEC} \simeq (\sqrt{2}-1)\frac{\sqrt{|Q|}}{\alpha}\,,
\end{equation}
so that increasing $\alpha$ systematically \emph{reduces} the EC-satisfying region.  Practically, when the outer horizon satisfies $r_h>r_{\rm NEC}$, the horizon probes a regime where $\rho+P_\theta<0$ and macroscopic NEC/WEC violations occur.  This result explains why large $\alpha$ produces ``exotic'' macroscopic behaviour in the spacetime solution: the enhanced $\alpha$-dependent terms in the stress tensor (dominant at order $\alpha^2,\alpha^4\log r_h,\dots$) provide the extra freedom necessary to violate classical ECs at horizon scales.  

    {We provide the explicit energy-condition radii: $r_{\rm NEC}=(\sqrt2-1)\sqrt{|Q|}/\alpha$ and $r_{\rm WEC}=((\sqrt{97}-7)/24)\sqrt{|Q|}/\alpha$. NEC holds only for $0<r<r_{\rm NEC}$​ (and WEC only inside $r_{\rm WEC}$), so an outer horizon with $r_h>r_{\rm NEC}$​ lies in an NEC-violating regime. Consequently, thermodynamic phenomena observed for parameter choices that place the outer horizon beyond these radii must be classified as occurring in an exotic (energy-condition-violating) regime; conversely, when $r_h<r_{\rm NEC}$, the thermodynamic features lie in an EC-respecting domain.}


\section{Thermodynamic topology}
\label{topology}
Recent developments in black hole thermodynamics highlight the role of thermodynamic topology in elucidating complex phase structures. This approach, rooted in topological techniques originally introduced by Duan for relativistic particle systems, treats black holes as thermodynamic defects. Within this framework, the zero points of a suitably defined vector field correspond to critical points associated with phase transitions. Each of these points carries an integer winding number, which serves to classify the system’s topology.

The overall shape of the generalized off-shell free energy was introduced in \cite{29,york}, defined as follows:
\begin{equation}
\mathcal{F} = M - \frac{S}{\tau}, \label{eq:free_energy}
\end{equation}
where $\tau$ is an inverse temperature parameter defining the thermodynamic ensemble. From $\mathcal{F}$, we define the two-component vector field \cite{29}
\begin{equation}
\boldsymbol{\varphi} = (\varphi_\tau,\,\varphi_\Theta) = \left(\frac{\partial F}{\partial S},\,-\cot\Theta\,\csc\Theta\right), \label{phi vector}
\end{equation}
with coordinates $(\tau, \Theta)$ in the extended parameter space. The zeros of $\boldsymbol{\phi}$ occur at
\begin{equation}
(\tau,\Theta) = \left(\frac{1}{T},\,\frac{\pi}{2}\right),
\end{equation}
where $T$ is the equilibrium Hawking temperature of the black hole in a heat bath.

We construct the unit vector $n^a = \varphi^a/\|\varphi\|$, satisfying $n^a n^a = 1$, and define the conserved topological current in three-dimensional parameter space $(t, S, \Theta)$ \cite{Duan:1998it,Wei:2021vdx, Sekhmani:2025zwc}:
\begin{equation}
j^\mu = \frac{1}{2\pi}\,\epsilon^{\mu\nu\rho}\,\epsilon_{ab}\,\partial_\nu n^a\,\partial_\rho n^b, \label{eq:topo_current}
\end{equation}
where $\epsilon^{\mu\nu\rho}$ and $\epsilon_{ab}$ are the Levi-Civita symbols in 3D and 2D, respectively. The projection of this current onto the $(\tau,\Theta)$ plane gives the topological density
\begin{equation}
j^0 = \delta^{(2)}(\boldsymbol{\varphi})\,J^0\left(\frac{\boldsymbol{\varphi}}{x}\right),
\end{equation}
where $J^0(\boldsymbol{\varphi}/x)$ is the Jacobian determinant of the vector field $\boldsymbol{\varphi}$. 

Integrating \( j^0 \) over a compact domain \( \Sigma \) in the  entropy-angle space, bounded by smooth contours parametrized as
\begin{equation}
S = S_1\cos\nu + S_0, \quad \Theta = S_2\sin\nu + \frac{\pi}{2}, \quad \nu \in (0,2\pi), \label{eq:contour}
\end{equation}
yields the total topological charge \cite{Wei:2021vdx}
\begin{equation}
W = \int_\Sigma j^0\,dS\,d\Theta = \sum_i w_i, \label{eq:total_charge}
\end{equation}
with each winding number
\begin{equation}
w_i = \frac{1}{2\pi} \oint_{C_i} \epsilon^{ab} n^a\,dn^b, \label{eq:winding}
\end{equation}
representing the net circulation of the vector field around a zero point (defect), which correlates with the nature of the thermodynamic phase.

It is worth noting that $S_1$ ($S_2$) in Eq. \eqref{eq:contour} represents the oscillation amplitude of $S$ ($\Theta$) around the central point $S_0$ ($\pi/2$). In other words, $S_1$ ($S_2$) governs the contour’s extent along the entropy (angular) direction. These parameters are not universal constants but are chosen in accordance with the requirements of the topological analysis. Specifically, $S_1$ and $S_2$ are selected small enough so that the closed contour defined in Eq.~\eqref{eq:contour} encloses only a single zero of the vector field $\phi$. This ensures an unambiguous identification of individual critical points and a consistent evaluation of the corresponding winding number. It is essential that the amplitudes do not cause the contour to intersect multiple singularities, as this would undermine the interpretation of the associated topological charge. Geometrically, $S_1$ and $S_2$ determine the extent of the integration loop along the entropy and angular directions, respectively, and must be adjusted to the local structure of the free-energy landscape to preserve the locality of the topological classification~\cite{Wei:2021vdx,Wei:2022dzw}.

Now, combining the definition~\eqref{eq:free_energy} with the expressions in Eqs.~\eqref{M} and~\eqref{SS}, one obtains the generalized off-shell free energy $\mathcal{F}$ in terms of the parameter space pertinent to the spacetime solution as
\begin{align}
		\mathcal{F}&=M-\frac{S}{\tau }\nonumber\\
        &\hspace{-4mm}=\frac{1}{6} \bigg\{r_h \left[-12 \alpha ^4 r_h^2 \log \left(\frac{\sqrt{| Q| }}{\alpha  r_h}+1\right)+8 \pi  P r_h^2-\frac{6 \pi  r_h}{\tau }+3\right]\nonumber\\
        &\hspace{-4mm}+4
   \alpha  | Q| ^{3/2}+12 \alpha ^3 r_h^2 \sqrt{| Q| }-3 \alpha ^2 r_h | Q| \bigg\}.
	\end{align}
The stationary condition at fixed $\tau$ is given by $\partial_S \mathcal{F}=\varphi^{S}=0$, implying that equilibrium configurations coincide with solutions of the thermal relation $\tau=1/T(r_h)$. To analyze the free-energy landscape, it is instructive to examine the small- and large-radius asymptotics. For $r_h\ll \sqrt{|Q|}/\alpha$, using $\ln(\tfrac{\sqrt{|Q|}}{\alpha r_h}+1)\simeq\ln(\tfrac{\sqrt{|Q|}}{\alpha r_h})$, one finds the expansion  \begin{equation}
\label{eq:F_small}
\mathcal{F}(r_h)=\frac{2}{3}\alpha |Q|^{3/2}+r_h\Big(\tfrac{1}{2}-\tfrac{\alpha^{2}|Q|}{2}\Big)+\mathcal{O}^,\!\big(r_h^{2}\ln(1/r_h)\big),
\end{equation}which entails the following results:
\begin{itemize}
\item The constant offset $\tfrac{2}{3}\alpha|Q|^{3/2}$ shows that the combined effect of $\alpha$ and $Q$ gives a finite free-energy bias at very small radii.
\item The linear coefficient changes sign when $\alpha^{2}|Q|=1$; hence, $\alpha^{2}|Q|=1$ is a small-radius threshold that controls whether $\mathcal{F}$ initially increases or decreases with $r_h$.
\end{itemize}
At the opposite extreme, for $r_h\gg\sqrt{|Q|}/\alpha$, using $\ln(1+\varepsilon)\simeq \varepsilon$ with $\varepsilon=\sqrt{|Q|}/(\alpha r_h)$, the free energy grows as  \begin{equation}
\label{eq:F_large}
\mathcal{F}(r_h)\sim \frac{4\pi}{3}P\,r_h^{3}, \qquad (r_h\to\infty),
\end{equation}
so that the pressure $P$ controls the energetic scale of large black holes and dictates the location and depth of large black hole (LBH) minima. Collecting these results, one sees that $\alpha$ and $Q$ dominantly regulate the small- and intermediate-radius branches: the threshold $\alpha^{2}|Q|=1$ marks a qualitative transition in small black hole (SBH) behaviour, while logarithmic contributions proportional to $\alpha^{4}$ generate slowly varying curvature in the intermediate regime, thereby introducing inflection points and enabling an intermediate black hole (IBH) branch. Meanwhile, $P$ governs the LBH sector through its cubic growth, and hence strongly influences the global free-energy competition among SBH, IBH, and LBH configurations.

\begin{figure}[t!]
\centering
\includegraphics[scale=0.79]{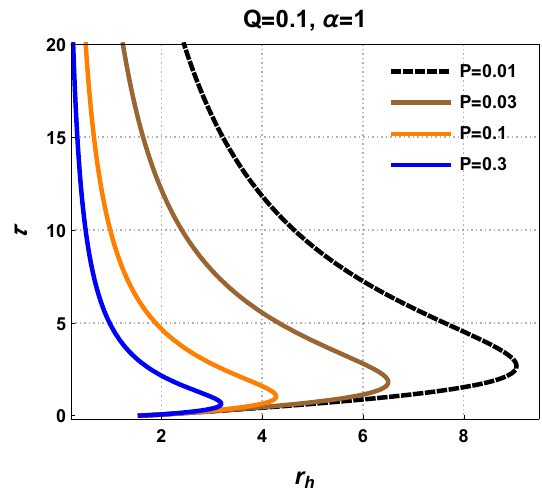}\\
\includegraphics[scale=0.82]{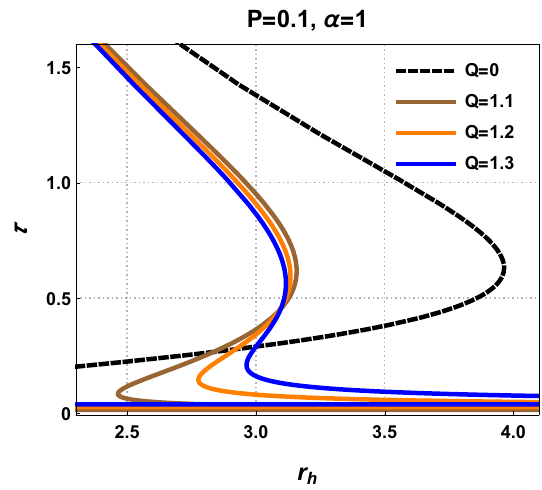}\\
\includegraphics[scale=0.82]{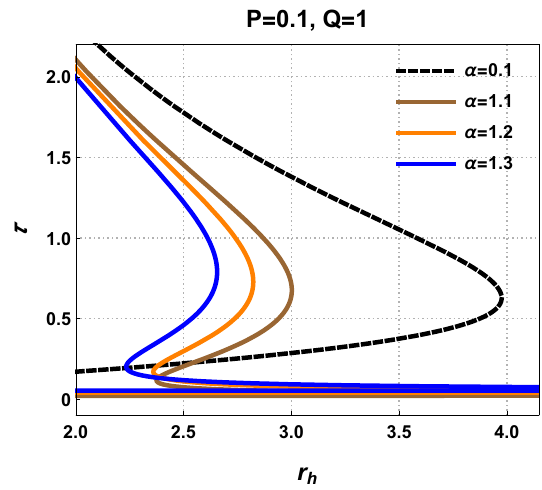}
\caption{Defect curve $\tau$ versus $r_h$ for various values of the pressure $P$, the electric charge $Q$ and the coupling parameter $\alpha$. }
\label{Fig4}
\end{figure}

	Using Eq. \eqref{phi vector}, components of the vector $\phi$ are found to be :
        \begin{align}
		\varphi^{S} &=-6 \alpha ^4 r_h^2 \log \left(\frac{\sqrt{| Q| }}{\alpha  r_h}+1\right)+4 \pi  P r_h^2-\frac{2 \pi  r_h}{\tau }+\frac{1}{2}\nonumber\\
        &+2 \alpha ^3 r_h \sqrt{| Q| } \left(\frac{\alpha 
   r_h}{\sqrt{| Q| }+\alpha  r_h}+2\right)-\frac{\alpha ^2 | Q| }{2}
  \end{align}
  and
	\begin{equation}
		\varphi ^{\Theta }=-\cot \Theta ~\csc \Theta.
\end{equation}
Because $\varphi^\Theta$ depends solely on $\Theta$ (we use $\varphi^\Theta=-\cot\Theta\csc\Theta$ evaluated at $\Theta=\pi/2$), the Duan Jacobian at a zero simplifies to
\begin{equation}
J_0=\partial_S \varphi^S\big|_{\text{zero}}=\partial_S T,
\end{equation}
and using standard thermodynamic identities $C\equiv T(\partial S/\partial T)\Rightarrow \partial_S T=T/C$, one finds
\begin{equation}
J_0=\frac{T}{C},
\end{equation}
so that the winding number associated with each zero is directly given by the sign of the heat capacity:
\begin{align}
w&=+1 \;\Longleftrightarrow\; C>0 \;(\text{locally stable}),\\
w&=-1 \;\Longleftrightarrow\; C<0 \;(\text{locally unstable}).
\end{align}

In practice, using $S=\pi r_h^2$ this translates into the slope-stability rule
\begin{equation}
C>0 \;\Longleftrightarrow\; \frac{d\tau}{dr_h}<0,\qquad
C<0 \;\Longleftrightarrow\; \frac{d\tau}{dr_h}>0,
\end{equation}
allowing one to read off local thermodynamic stability directly from $\tau(r_h)$ plots: a falling branch corresponds to a stable ($w=+1$) equilibrium, while a rising branch is unstable ($w=-1$).

The small-radius expansion of the Duan thermodynamic vector yields the leading term
\begin{equation}
\label{rob}
\varphi^S(0^+) = \frac{1}{2}-\frac{1}{2}\alpha^2|Q| + \mathcal{O}(r_h).
\end{equation}
This leads to a compact topological dictionary: if $\varphi^S(0^+)>0$ (subcritical), the small black hole (SBH) branch emerges with $w=-1$, and the phase structure is typically governed by first-order SBH/LBH competition. At $\varphi^S(0^+)=0$ (critical), one finds $J_0 \to 0$ and $C \to \infty$ at the tangency point, enabling the pair creation or annihilation of topological defects and realizing genuine second-order criticality. For $\varphi^S(0^+)<0$ (supercritical), the SBH branch emerges with $w=+1$, an intermediate branch (IBH) is generically allowed, and vertical tangencies in $\tau(r_h)$ lead to continuous transitions.  Thus, the dimensionless combination $\alpha^2 |Q|$ serves as a sharp analytic classifier:
\begin{itemize}
    \item subcritical ($\alpha^2|Q|<1$), \item critical ($\alpha^2|Q|=1$), \item supercritical ($\alpha^2|Q|>1$).
\end{itemize}

Subsequently, we identify the zero points or singularities of the vector field. A zero point always occurs at $\Theta = \tfrac{\pi}{2}$, owing to the specific choice of the $\Theta$-component of the field. To locate additional zero points, we determine an equation for $\tau$ by solving $\varphi^{S}=0$, which takes the form:
    \begin{widetext}
        \begin{equation}
	\tau=\frac{4 \pi  r_h \left(\sqrt{| Q| }+\alpha  r_h\right)}{\sqrt{| Q| } \left[-\alpha ^2 | Q| +7 \alpha ^3 r_h \sqrt{| Q| }+4 r_h^2 \left(3 \alpha ^4+2 \pi 
   P\right)+1\right]-12 \alpha ^4 r_h^2 \left(\sqrt{| Q| }+\alpha  r_h\right) \log \left(\frac{\sqrt{| Q| }}{\alpha  r_h}+1\right)+\alpha  r_h \left(8
   \pi  P r_h^2+1\right)}
	\end{equation}
    \end{widetext}
    Expanding the denominator of $\tau(r_h)$ for $r_h\to 0$ gives the leading behaviour 
\begin{equation}
\label{eq:tau_small}
\tau(r_h)\simeq \frac{4\pi r_h}{1-\alpha^{2}|Q|}+\mathcal{O}(r_h^{2}),
\end{equation}
which makes clear that the sign of $1-\alpha^{2}|Q|$ dictates the stability of the SBH branch. If $1-\alpha^{2}|Q|>0$, then $\tau\propto +r_h$ near the origin, implying $d\tau/dr_h>0$ and thus thermodynamic instability ($C<0$, $w=-1$); conversely, if $1-\alpha^{2}|Q|<0$, then $\tau\propto -r_h$, so $d\tau/dr_h<0$ and the branch is thermodynamically stable ($C>0$, $w=+1$). Hence, the dimensionless combination $\alpha^{2}|Q|$ acts as a control parameter that flips SBH stability across the threshold $\alpha^{2}|Q|=1$. At the opposite limit, for $r_h\to\infty$, the dominant terms yield 
\begin{equation}
\label{eq:tau_large}
\tau(r_h)\sim \frac{1}{2Pr_h}, \qquad \Longrightarrow \qquad T(r_h)\sim 2Pr_h,
\end{equation}
demonstrating that the pressure $P$ sets the temperature scale of large black holes and shifts LBH equilibria to different $\tau$ values. 

Collecting these observations, one can see that
\begin{itemize}
\item $Q$ primarily affects the small and intermediate regimes, entering through terms such as $\sqrt{|Q|}$ and $|Q|^{3/2}$. Increasing $Q$ can drive the SBH across the threshold $\alpha^{2}|Q|=1$, thereby flipping its stability.
\item $\alpha$ governs the structure of the free energy through higher-order contributions ($\alpha^{2}, \alpha^{3}, \alpha^{4}$) and logarithmic terms. It controls the small-radius threshold and amplifies the curvature effects responsible for creating or removing an IBH.
\item $P$ controls the large-radius sector through the cubic growth of $\mathcal{F}$, setting the LBH temperature scale and dominating the global free-energy competition among the SBH, IBH and LBH phases.
\end{itemize}

The plots in Fig. \ref{Fig4} display the equilibrium locus $\tau(r_h)=1/T(r_h)$ and therefore contains all candidate black hole phases: Each intersection of a horizontal line $\tau=const$ with the plotted curve is a thermodynamic equilibrium (a zero of $\varphi^{S}$). Two analytic observations make this plot especially diagnostic: $\varphi^{S}=\partial_{S}\mathcal{F}=T-1/\tau$, so equilibria satisfy $\tau=1/T(r_h)$, and the Jacobian at a zero reduces to the heat-capacity combination $J_{0}=\partial_{S}\varphi^{S}=\partial_{S}T=T/C$. Hence, the slope rule follows immediately: a falling segment of $\tau(r_h)$ (i.e., $d\tau/dr_h<0$) corresponds to $C>0$ and carries winding $w=+1$ (locally stable), while a rising segment ($d\tau/dr_h>0$) corresponds to $C<0$ and $w=-1$ (locally unstable). The small-radius expansion shows the compact threshold $\alpha^{2}|Q|=1$, which controls the small black hole stability flip: for $\alpha^{2}|Q|<1$ the small-radius branch begins with $d\tau/dr_h>0$ (SBH unstable), whereas for $\alpha^{2}|Q|>1$ the initial slope is negative and the SBH becomes locally stable. 

Physically, $Q$ (through $\sqrt{|Q|}$ and $|Q|^{3/2}$) primarily sculpts the small/intermediate structure, $\alpha$ (through high powers and the $\alpha^{4}\log$ term) is the main shape controller that creates/removes inflection points (thus enabling or suppressing the intermediate black hole phase), and $P$ governs the large-radius sector (since $\mathcal{F}\sim4/3\pi Pr_h^{3}$ and $\tau\sim 1/2Pr_h$), shifting the LBH branch vertically. 

Finally, topological pair creation/annihilation events occur exactly at tangency points $d\tau/dr_h=0$ (where $\partial_{S}T=0$ and $C\to\infty$). In addition, the small-radius stability is controlled by the dimensionless threshold $\alpha^{2}|Q|=1$: for $\alpha^{2}|Q|<1$ (e.g., $\alpha=0.8,\;Q=1$) the $\tau(r_h)$ curve rises near $r_h=0$ so the SBH is locally unstable, whereas for $\alpha^{2}|Q|>1$ (e.g. $\alpha=1.2,\;Q=1$) it falls and the SBH becomes locally stable; the marginal case $\alpha^{2}|Q|=1$ (e.g. $\alpha=1,\;Q=1$) marks where a tangency can produce a second-order phase transition ($d\tau/dr_h=0$, $C\to\infty$). The pressure $P$ acts mainly at large radii: increasing $P$ (for example $P=0.05\to0.2$) shifts the LBH branch to lower $\tau$ (higher $T$) and typically narrows the $\tau$-defect curve where three branches coexist.

\begin{figure}[t!]
\begin{center}
\subfigure[~Defect curve using $\alpha=2,\,P=0.1$ and $Q=1$.]  {\label{Parameter_Space_RN_AdSd}
\includegraphics[height=5.5cm,width=7cm]{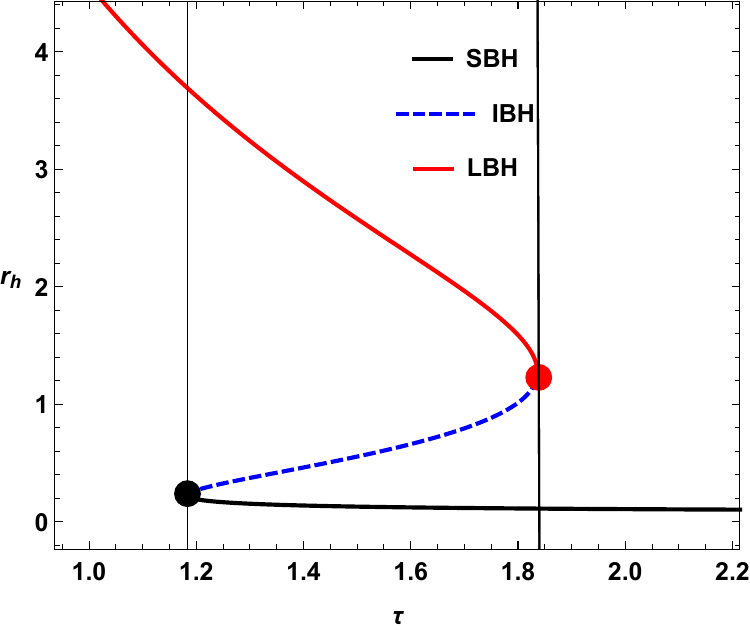}}\hfill
\subfigure[~Vector field]  {\label{Parameter_Space_RN_dSe}
\includegraphics[height=5.5cm,width=7.2cm]{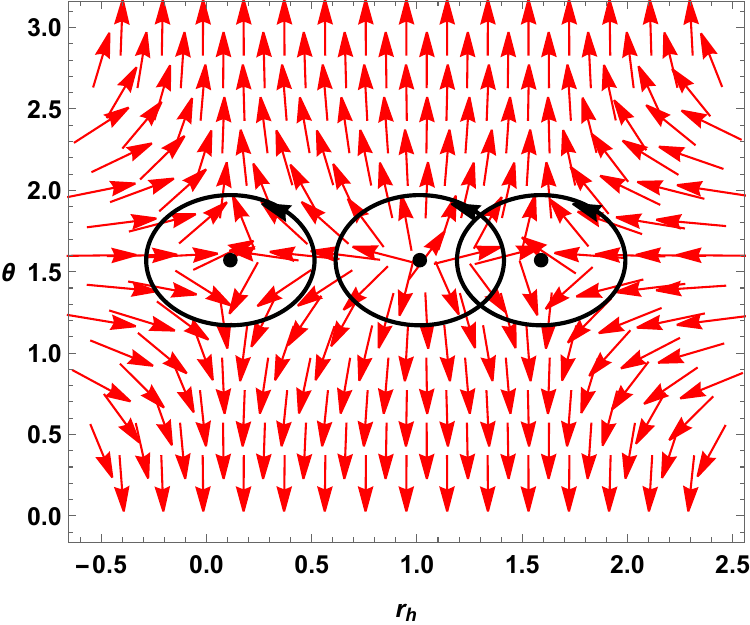}}\hfill
\subfigure[~Winding numbers]  {\label{Parameter_Space_RN_dSf}
\includegraphics[height=5.5cm,width=7.5cm]{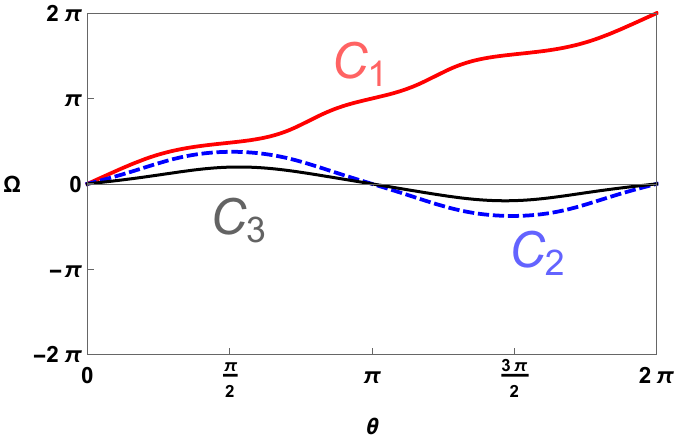}}
\end{center}
\caption{Thermodynamic topology for RegMax-AdS BHs using $Q=1,\,P=0.1,\alpha=2$ and $\tau=2.5$.}
\label{Fig5}
\end{figure}

For the supercritical coupling $\alpha=2$ (see Fig. \ref{Fig5}), the defect curve $\tau(r_h)$ and the associated Duan vector field (panels (a)–(c)) show the characteristic three-branch structure (small-, intermediate- and large-black hole) and a topological pair $(\pm1)$ pattern. This is a direct consequence of the small-radius threshold being passed: $\alpha^2|Q| = 4 \gg 1$ (for $Q = 1$), so the small-r expansion of $\phi^S$ is negative and the SBH branch emerges as thermodynamically stable $(C > 0)$. The presence of an intermediate branch and the vertical tangency of $\tau(r_h) (d\tau/dr_h = 0)$ mark locations where the heat capacity diverges and a second-order (continuous) critical behaviour (pair creation/annihilation of topological defects) can occur. The winding assignments shown in panel (c) therefore track stability: the SBH carries $w = +1$ (locally stable), the IBH carries $w = −1$ (unstable), and the LBH returns to $w = +1$, consistent with the slope rule and the sign of $C$.

\begin{figure}[t!]
\begin{center}
\subfigure[~Defect curve]  {\label{Parameter_Space_RN_AdSd2}
\includegraphics[height=5.5cm,width=7cm]{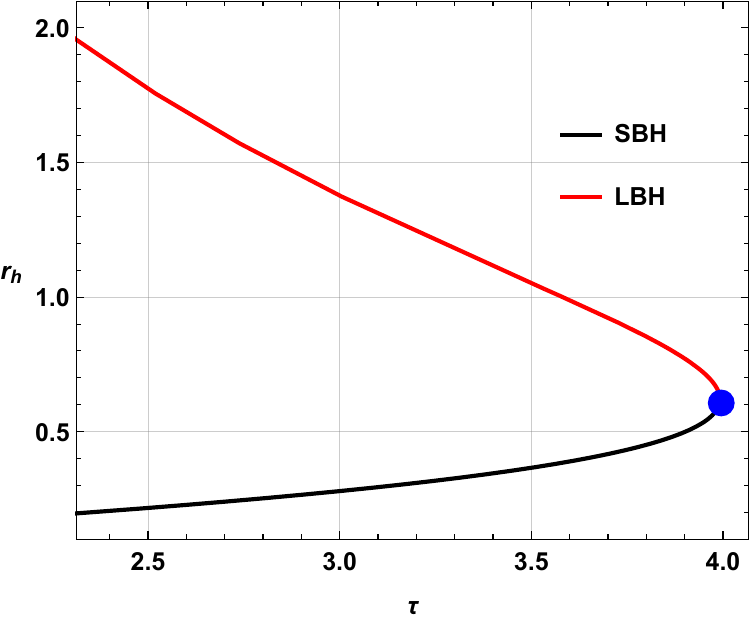}}\hfill
\subfigure[~Vector field]  {\label{Parameter_Space_RN_dSe2}
\includegraphics[height=5.5cm,width=7.2cm]{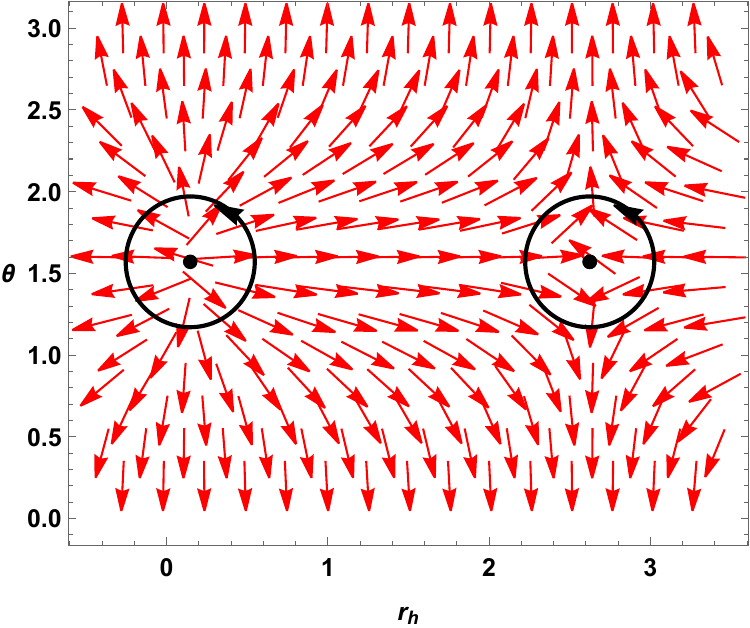}}\hfill
\subfigure[~Winding numbers]  {\label{Parameter_Space_RN_dSf2}
\includegraphics[height=5.5cm,width=7.5cm]{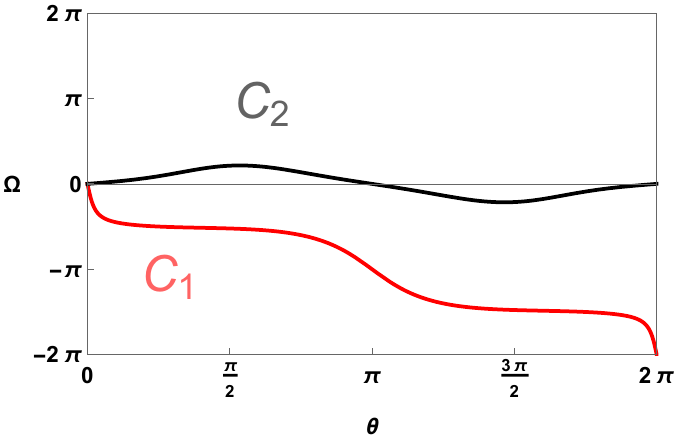}}
\end{center}
\caption{Thermodynamic topology for RegMax-AdS BHs using $Q=1,\,P=0.1,\,\alpha=0.2$ and $\tau=2.5$.}
\label{Fig6}
\end{figure}

By contrast, for the subcritical coupling $\alpha= 0.2$, the defect curve collapses to the simpler topology displayed in Fig. \ref{Fig6}. The small radius threshold $\alpha^2|Q| = 0.04 \ll 1$ places the SBH branch in the thermodynamically unstable sector $(C < 0)$ at small $r_h$, and the system does not develop the IBH region that produces a continuous second-order critical transition. Instead, the dominant transition pattern is a small/large first-order (discontinuous) coexistence (latent-heat style) or a direct crossover determined by free-energy competition between SBH and LBH. 

Topologically, fewer zeros/defects appear in the $(\tau,\theta)$ plane, and the winding number accounting shows a net simplification (fewer $\pm1$ pairs). This explains why the same $P$ and $Q$ produce qualitatively different phase diagrams when $\alpha$ is reduced from $2$ to $0.2$.

\section{Thermal geometry}
\label{Sectthermalgeometry}

Since a well-defined temperature can characterize black holes, it is natural to speculate that they may also exhibit an underlying microscopic structure. In recent years, significant effort has been devoted to uncovering the possible constituents and interactions responsible for such a structure~\cite{Cai:1998ep,Wei:2015iwa,Wei:2019uqg,Guo:2019oad,Xu:2020gud,Ghosh:2020kba,Xu:2020ftx,Prom,Dehghani:2023yph,Luciano:2023fyr,Luciano:2023bai,2025PhyB..71417484S,Ghaffari:2023vcw,Sekhmani:2024udl}. The findings of these investigations point toward the intriguing possibility that black holes behave as if they are composed of microscopic degrees of freedom, in a manner reminiscent of molecular interactions within a non-ideal fluid.

A common method for investigating the interactions among the conjectured microscopic constituents of black holes is through thermodynamic geometry. This approach assigns a geometric structure to the macroscopic thermodynamic phase space, allowing one to encode features of the underlying statistical mechanics. Notably, the geometric formulations introduced independently by Weinhold~\cite{Wein1} and by Ruppeiner~\cite{Rupp1,Rupp2} have been shown to provide valuable insight into the qualitative nature of microscopic interactions in conventional thermodynamic systems. Further developments along these lines have subsequently been carried out in~\cite{Quevedo:2006xk,Quevedo:2007mj}.

The scalar curvature derived from the thermodynamic metric serves as a diagnostic tool for identifying the prevailing type of microscopic interaction. A negative curvature is generally associated with attractive interactions among the constituents, whereas a positive curvature reflects repulsive behavior. When the curvature vanishes, it either signifies the absence of interactions, as in the case of an ideal gas, or represents an exact cancellation between attractive and repulsive contributions.

To investigate how our gravity framework influences the 
geometrothermodynamic properties of black holes, we proceed to evaluate 
the Weinhold and Ruppeiner scalar curvatures using the thermodynamic 
relations established in Sec.~\ref{sec:bhsolution}. In this framework, the Weinhold 
metric is introduced as the Hessian of the internal energy with respect 
to the selected thermodynamic variables~\cite{Wein1}. {For black hole 
systems, where the internal energy is identified with the mass 
(see~Eq.~\eqref{M}), the corresponding metric components take the form:
$g_{ij}^W = \partial_i \partial_j M(S, p, q) \Longrightarrow  ds^2_W = g^W_{ij} dx^i dx^j$,
where \( x^i \) denotes a set of independent thermodynamic fluctuation coordinates.}

{On the other hand, within the Ruppeiner approach, the entropy is regarded as the fundamental 
thermodynamic potential. The corresponding metric is therefore obtained 
from the negative Hessian of $S$ with respect to the thermodynamic
variables, namely
\begin{equation}
\label{R1}
g_{ij}^{\text{Rup}} = -\,\partial_i \partial_j S \, .
\end{equation}
Combining the Weinhold metric with Eq. \eqref{R1} and the definition of the 
temperature, one finds that the Weinhold and Ruppeiner metrics are linked 
through a conformal transformation in which the temperature plays the role 
of the conformal factor, i.e.
\begin{equation}
\label{R2}
ds^{2}_{R} = \frac{1}{T}\, ds^{2}_{W}\, .
\end{equation}}

{Using the above tools, we now proceed to the geometrothermodynamic investigation 
of black holes in the RegMax theory. 
We resort to the Ruppeiner construction in the energy representation of thermodynamic geometry, owing to its well-known connection with 
fluctuation theory in statistical mechanics~\cite{Rupp1,Rupp2}. 
The analysis is performed in the canonical ensemble (fixed charge), 
where the entropy $S$ and the pressure $P$ are taken as the relevant 
thermodynamic fluctuation variables. 
Accordingly, the components of the Ruppeiner metric are defined as
\begin{equation}
g^{\text{Rup}}_{ij} = \frac{1}{T}\,
\frac{\partial^2 M(S,P)}{\partial x^i \partial x^j},
\qquad x^i = (S, P).
\label{RuppMetric}
\end{equation}
To streamline the notation, let us rewrite the Ruppeiner metric as
\begin{equation}
g^{\text{Rup}}=\begin{pmatrix}
a(x^1,x^2) & b(x^1,x^2) \\[2mm]
b(x^1,x^2) & 0
\end{pmatrix}\,,
\end{equation}
where $a(x^1,x^2)\equiv\frac{1}{T}\,
\frac{\partial^2 M}{\partial (x^1)^2}$ and $b(x^1,x^2)\equiv\frac{1}{T}\,
\frac{\partial^2 M}{\partial x^1 \partial x^2}=\frac{1}{T}\,
\frac{\partial^2 M}{\partial x^2 \partial x^1}$, while the element $g_{22}^{\text{Rup}}=0$, since the mass $M$ in Eq. \eqref{M} is linear in $P$. }

{The explicit computation of the scalar curvature of $g^{\text{Rup}}$ then gives
\begin{equation}
    {R}^{\text{Rup}}=\frac{1}{b^2}\left[\frac{\partial^2 a}{\partial (x^2)^2} - 2\frac{\partial^2 b}{\partial x^1\partial x^2}  \right]-\frac{1}{b^3}\left(\frac{\partial a}{\partial x^2} -2\frac{\partial b}{\partial x^1} \right)\frac{\partial b}{\partial x^2} 
    \,.
\end{equation}
By substituting the original thermodynamic variables, we obtain
\begin{eqnarray}
\label{Rscalar}
\nonumber
&&\hspace{-5mm}\mathcal{R}^{Rup} (r,p)=-\frac{\alpha\left[\alpha\left(2Q^2- r^2\right)-2\sqrt{Q}r\right]-Q}{\pi r^2\left(\sqrt{Q}+r\alpha\right)}\times\\[2mm]
\nonumber
&&\hspace{-4mm}\left\{r\alpha\left(1+8\pi P r^2\right)-12r^2\alpha^4(\sqrt{Q}+r\alpha)\log\left(1+\frac{\sqrt{Q}}{r\alpha}\right)\right.
\\[2mm]
&&\hspace{-4mm}+\left.\sqrt{Q}\left[1+4r^2(2\pi P+3\alpha^4)\right]-2 Q^{3/2}\alpha^2+6Qr\alpha^3
\right\}^{-1},
\end{eqnarray}
where, without loss of generality, we have assumed $Q > 0$.}



\begin{figure}[t!]
\centering
\includegraphics[width=7.7cm]{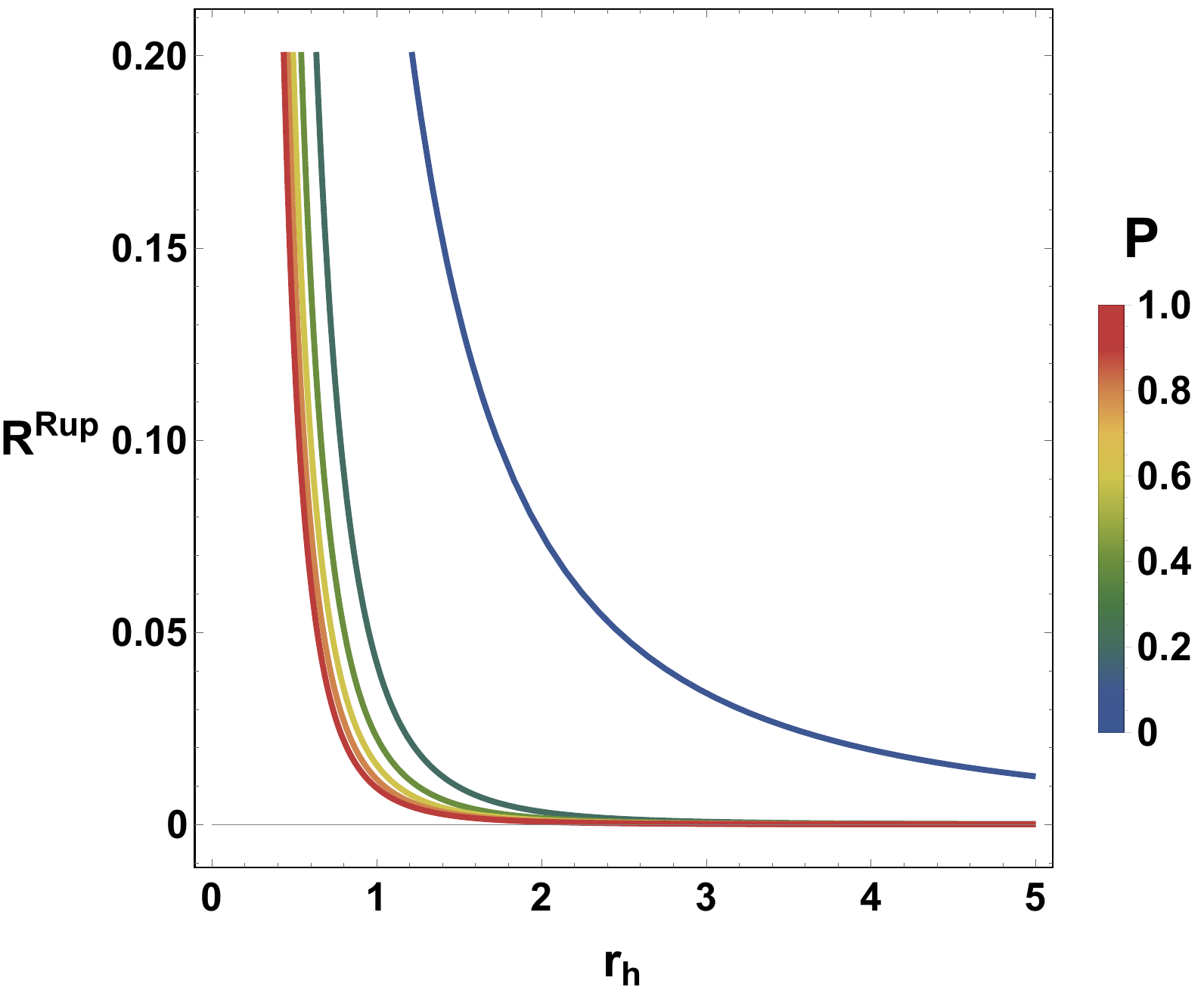}\\[2mm]
\includegraphics[width=8cm]{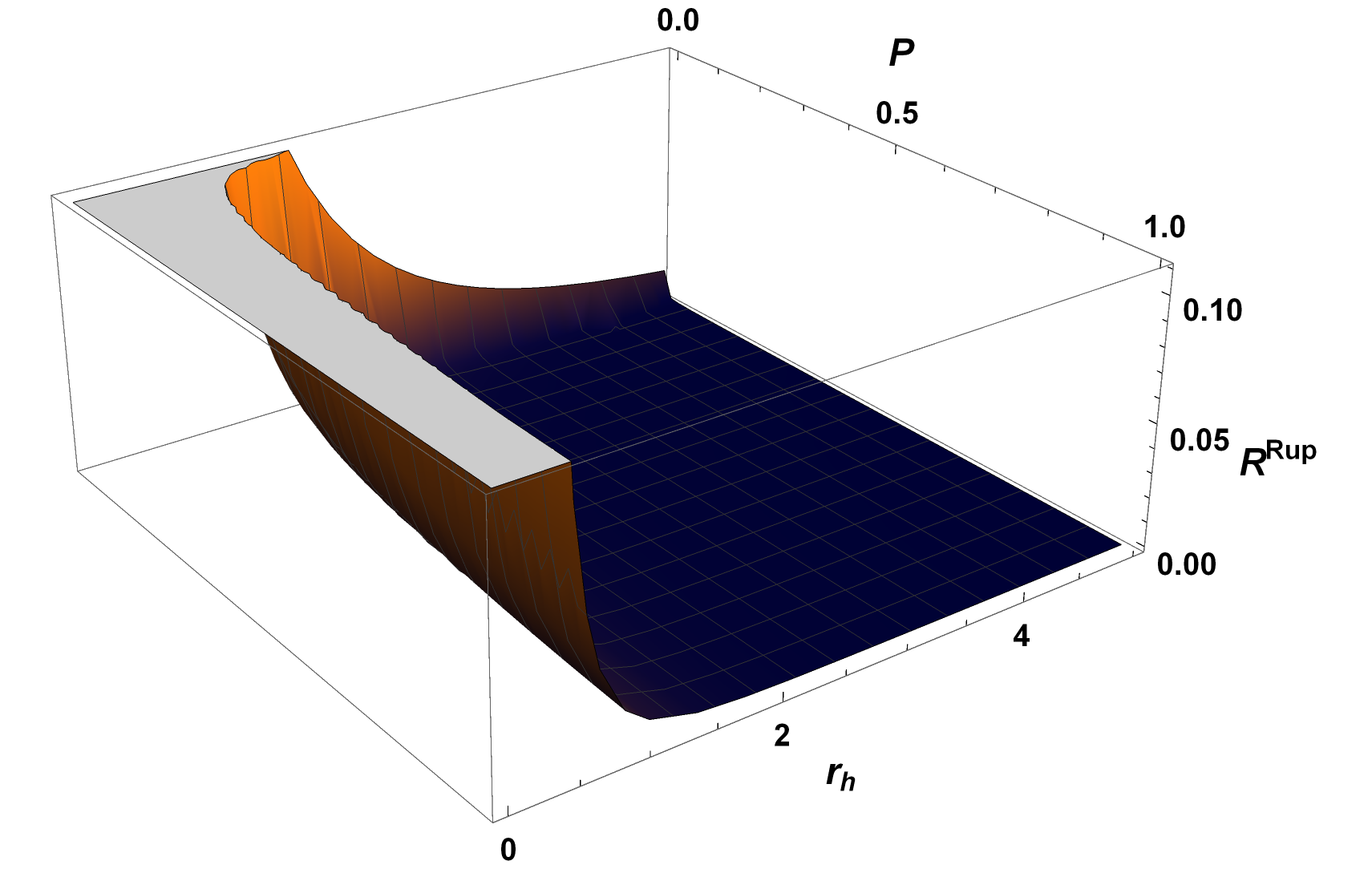}
\caption{Plots of $\mathcal{R}^{Rup}$ as a function of $r_h$ (upper panel) and $(r_h, P)$ (lower panel) in the subcritical regime ($\alpha=0.5$). We set the reference value $Q=1$.}
\label{SubRup1}
\end{figure}

\begin{figure}[t!]
\centering
\includegraphics[width=7.3cm]{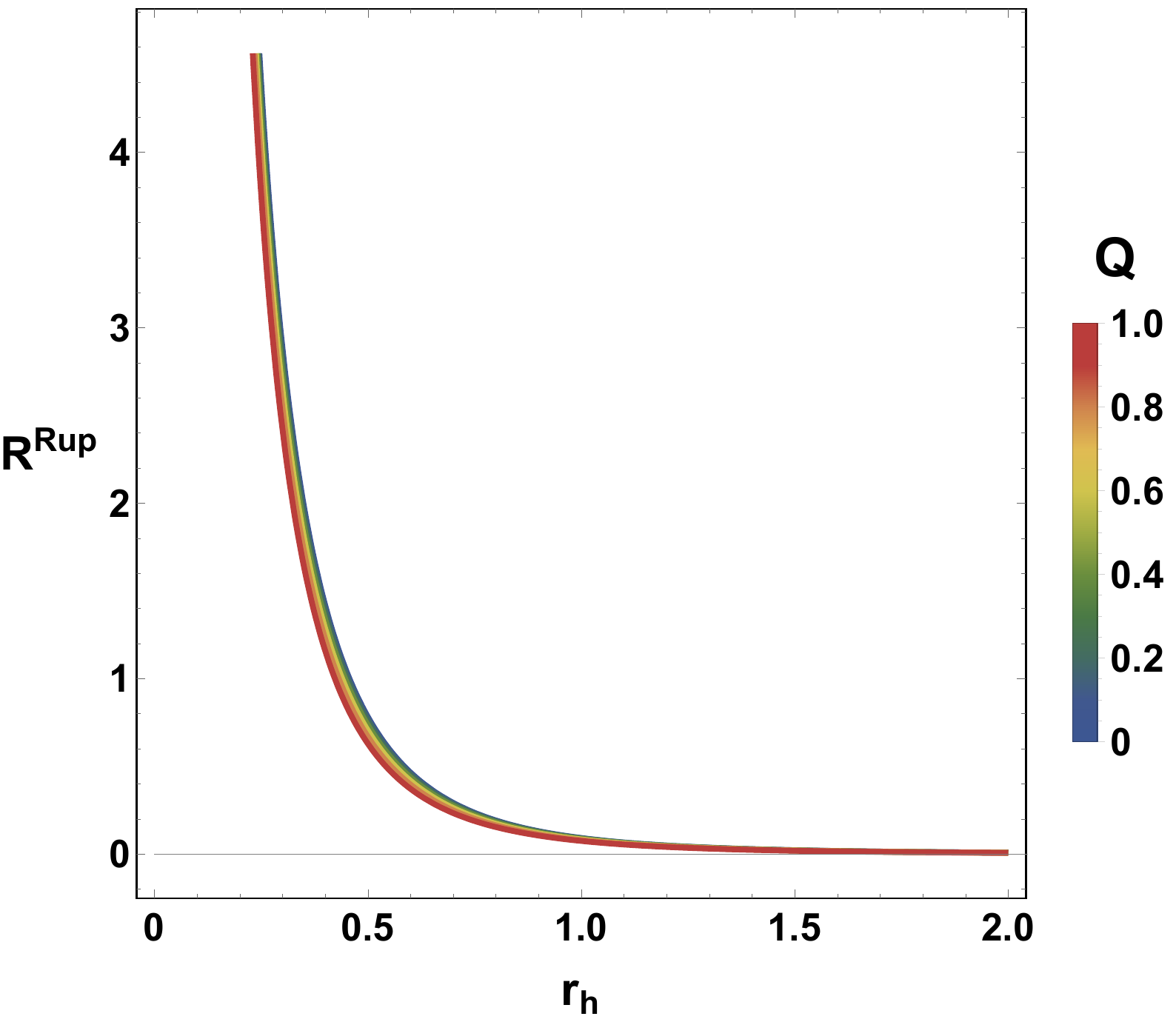}\\[2mm]
\includegraphics[width=8cm]{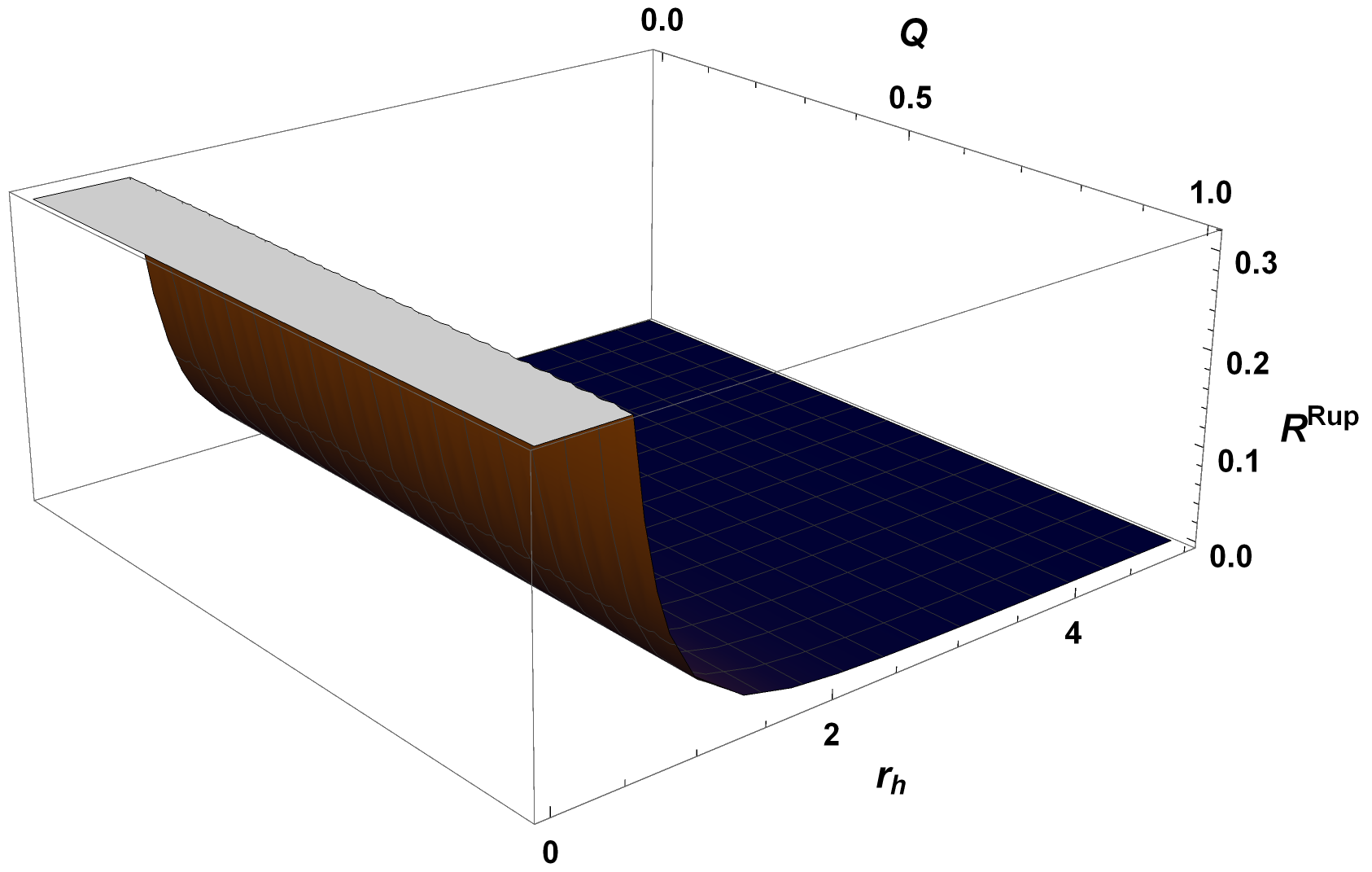}
\caption{Plots of $\mathcal{R}^{Rup}$ as a function of $r_h$ (upper panel) and $(r_h, Q)$ (lower panel) in the subcritical regime ($\alpha=0.5$). We set the reference value $P=0.1$.}
\label{SubRup2}
\end{figure}

\begin{figure}[t!]
\centering
\includegraphics[width=7.7cm]{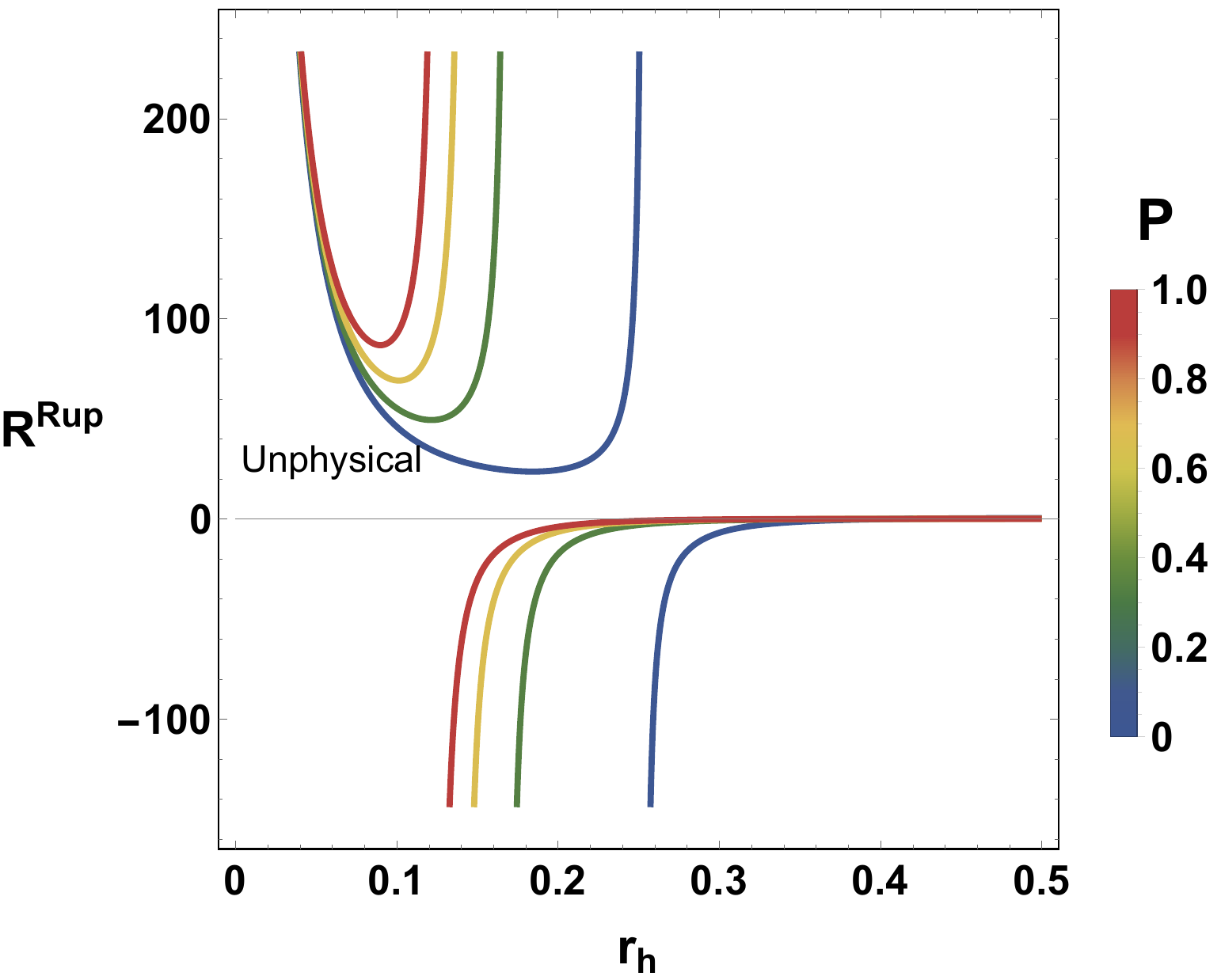}\\
\includegraphics[width=8cm]{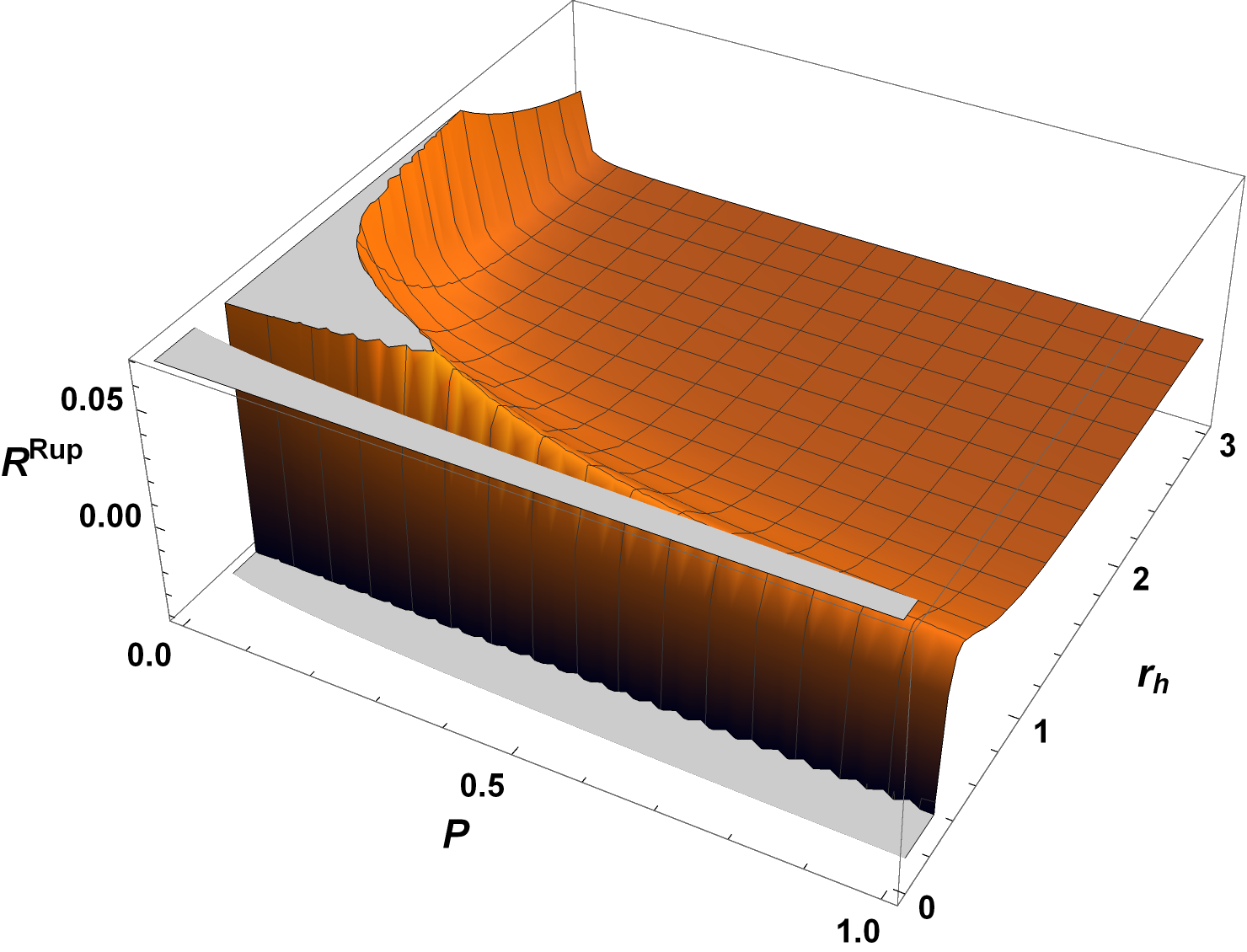}
\caption{Plots of $\mathcal{R}^{Rup}$ as a function of $r_h$ (upper panel) and $(r_h, P)$ (lower panel) in the critical regime ($\alpha=1$). We set the reference value $Q=1$.}
\label{CritRup}
\end{figure}

\begin{figure}[t!]
\centering
\includegraphics[width=7.7cm]{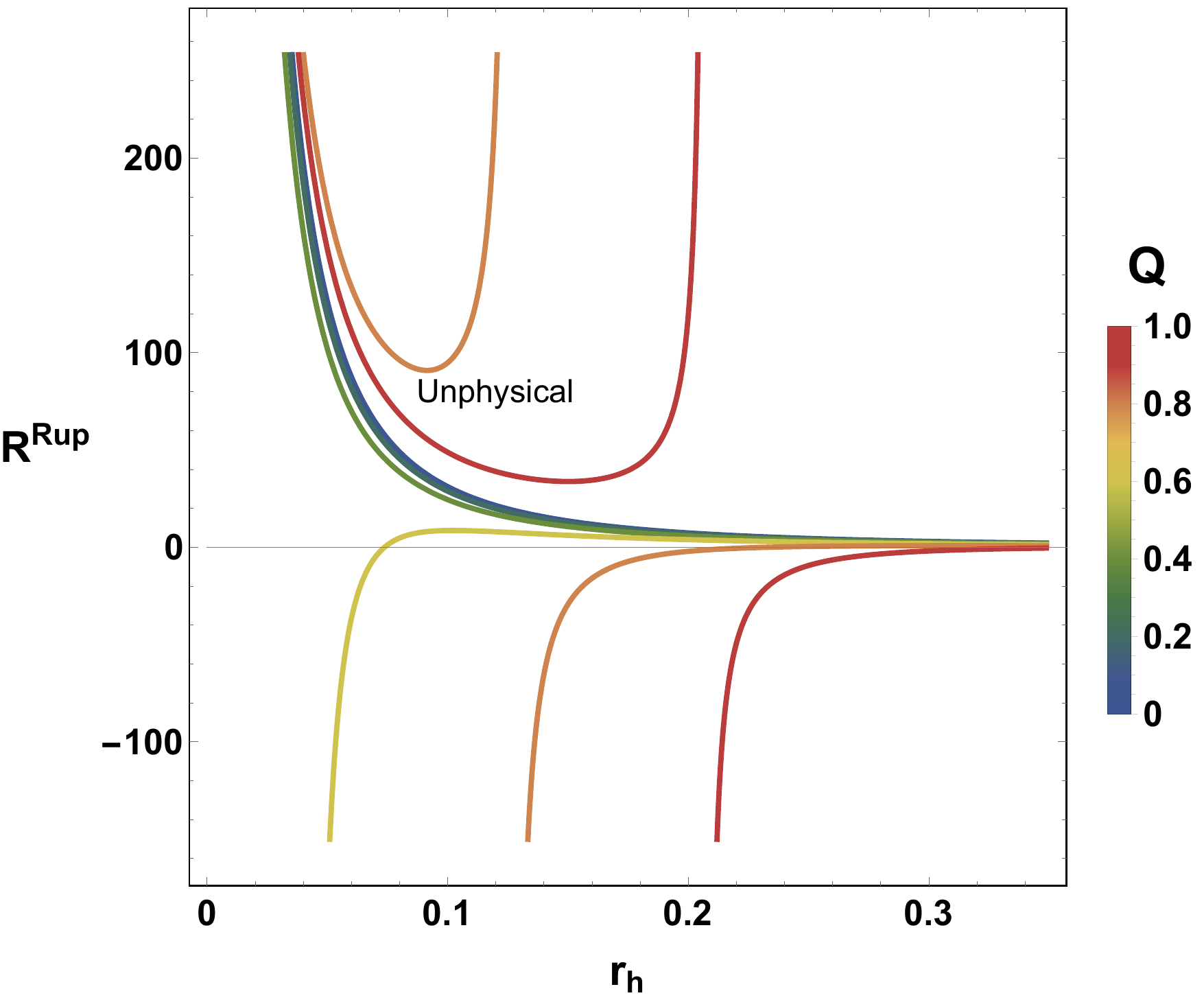}\\
\includegraphics[width=8cm]{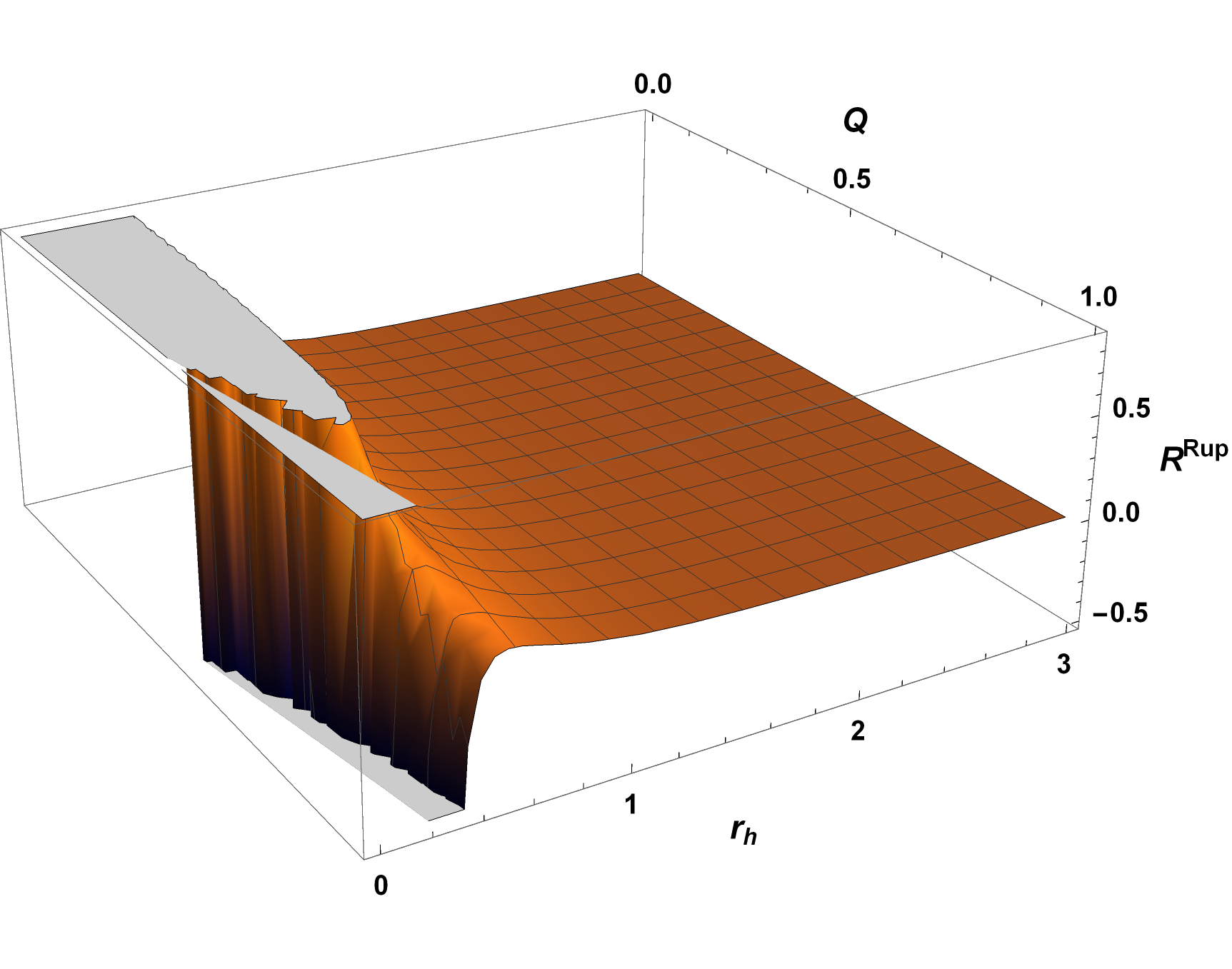}
\caption{Plots of $\mathcal{R}^{Rup}$ as a function of $r_h$ (upper panel) and $(r_h, P)$ (lower panel) in the supercritical regime ($\alpha=2$). We set the reference value $Q=1$.}
\label{SupCrit}
\end{figure}

The behavior of $\mathcal{R}^{Rup}$ is illustrated in Figs. \ref{SubRup1}-\ref{SubRup2} for the subcritical regime, in Fig. \ref{CritRup} for the critical regime, and in Fig. \ref{SupCrit} for the supercritical  regime, for selected values of the phase-space parameters. 
In particular, from Figs.~\ref{SubRup1}-\ref{SubRup2}, one clearly sees that the Ruppeiner curvature $R^{\mathrm{Rup}}$ takes positive values, which reveals the predominance of repulsive microscopic interactions. The figures further illustrate a systematic trend: for larger black holes, the strength of these interactions progressively weakens, and in the asymptotic regime the curvature tends to vanish. This indicates that, in the macroscopic limit, the microconstituents interact only very weakly, approaching an almost ideal thermodynamic behavior. 

Conversely, as the black hole shrinks, the  $R^{\mathrm{Rup}}$ grows steadily, signaling that the interactions become increasingly intense in the small-size regime. 
A qualitatively similar behavior can be observed by letting both the pressure $P$ and the charge $Q$ vary (see the 3D plots in Figs.~\ref{SubRup1}-\ref{SubRup2}).

Within the framework of geometrothermodynamics, this specific pattern originates from the way the thermodynamic metric encodes the response of the system to fluctuations. At large radii, the thermodynamic potentials vary only mildly with respect to the extensive parameters, so the corresponding susceptibilities remain nearly constant and the associated equilibrium manifold is only weakly curved. This explains why the curvature approaches zero in the large-radius limit. In contrast, when the black hole shrinks, the variations of the thermodynamic variables with respect to the state parameters become sharper, which enhances the curvature of the equilibrium manifold. The growth of the curvature is thus the geometric imprint of increasingly strong correlations in this regime.


{A richer phenomenology is observed in the critical regime (see Fig.~\ref{CritRup}). In this case, one can see that $\mathcal{R}^{Rup}$ exhibits an apparent singularity at a critical value of $r_h$, which depends parametrically on the pressure $P$. This singularity, however, is unphysical, since it corresponds to the point in phase space where the absolute temperature vanishes. Clearly, the physically meaningful region is that with $T>0$ and, thus, for values of $r_h$ strictly greater than the critical one. Therefore, such apparent singularity does not signal any genuine phase transition or critical behavior, but rather marks the boundary of the physically accessible region where the temperature vanishes.}

In contrast to the subcritical regime discussed above, within this physical region the curvature 
$\mathcal{R}^{\mathrm{Rup}}$ takes negative values for $r_h$ small enough, signaling 
attractive interactions among the microscopic constituents of the black hole. 
As $r_h$ increases, however, $\mathcal{R}^{\mathrm{Rup}}$ becomes positive, 
indicating that repulsive interactions dominate, and it then approaches zero 
asymptotically.


Finally, the supercritical regime is analyzed in Fig.~\ref{SupCrit}, which displays a behavior qualitatively similar to that of the critical regime. 

Summarizing the above analysis, we have found that the geometrothermodynamic properties of the black hole model under consideration exhibit a remarkably rich and nontrivial phenomenology. The Ruppeiner curvature $\mathcal{R}^{\mathrm{Rup}}$ encodes subtle information on the microscopic interactions, unveiling a clear interplay between the parameters $(r_h, P, Q)$ and, most importantly, the coupling constant $\alpha$. 
Altogether, the results we have obtained highlight the crucial role of the parameter $\alpha$ in shaping the phase-space structure of the equilibrium manifold, and show how different microscopic interaction patterns emerge depending on the coupling regime. The model thus provides a concrete example of how geometrothermodynamics can capture the nontrivial interplay of macroscopic and microscopic parameters, offering new insights into the effective statistical interactions governing black hole thermodynamics.

\subsection{Topological charge and Ruppeiner Ricci scalar relation}
We elaborate further on the connection between the topological charge and the Ruppeiner Ricci scalar. After restricting to the branch on the shell $r=r(\alpha,Q)$, we define the components of the function $\varphi^{S}(\alpha,Q)=T(\alpha,Q)$ and $\varphi^{\Theta}(\alpha,Q)=\Psi(\alpha,Q)$, as well as its Jacobian 
\begin{equation}
J(\alpha,Q)=\det\!\left(\frac{\partial \varphi^{b}}{\partial X^{a}}\right)
=\det\!\left(\frac{\partial^{2} M}{\partial X^{a}\partial X^{b}}\right)\,,
\end{equation}
where $T(\alpha,Q)$ and $\Psi(\alpha,Q)$ are the on-shell temperature and electric potential, respectively. By using the expression of $\mathcal{R}^{Rup}$ in Eq.~(\ref{Rscalar}), one then has the local identity 
\begin{equation}
\mathcal{R}^{Rup}(\alpha,Q)\;\sim\;\frac{N(\alpha,Q)\,T(\alpha,Q)^{4}}{[J(\alpha,Q)]^{2}},
\end{equation}
where $N(\alpha,Q)$ denotes a smooth numerator constructed from higher-order derivatives of the mass function $M(\alpha,Q)$. 
In other words, $N(\alpha,Q)$ encapsulates regular contributions from the Hessian and mixed partial derivatives of $M$, ensuring finiteness across the parameter space. 
This highlights that both geometric diagnostics, such as the Ruppeiner Ricci scalar, and topological diagnostics, via the Duan current, are ultimately governed by the same underlying Hessian structure.

As a result, the $\alpha$ parameter space splits into three distinct regions: 
\emph{(i)} \emph{subcritical} ($J(\alpha,Q)\neq 0$ throughout the entire surveyed $(\alpha,Q)$ domain), where $\det g \propto J/T^{2}$ remains finite, $\mathcal{R}^{Rup}$ is regular, and the $\varphi$–map contains no distinct integer defects (a topologically trivial phase); \emph{(ii)} \emph{critical} (an isolated point $(\alpha_{c},Q_{c})$ with $J(\alpha_{c},Q_{c})=0$, equivalently $D(\alpha_{c},Q_{c})=0$ in Eq.~(\ref{Rscalar})), where the Hessian collapses, linear responses (e.g., $C_{Q}$) diverge generically, and $|\mathcal{R}^{Rup}|\to\infty$ unless the numerator $N(\alpha_{c},Q_{c})$ vanishes at the same order. The point $(\alpha_{c},Q_{c})$ is therefore an isolated topological defect, with Duan index determined by the sign and multiplicity of the Jacobian around zero;
\emph{(iii)} \emph{supercritical} (parameters beyond the critical locus), where $J$ typically becomes nonzero again, $\mathcal{R}^{Rup}$ remains finite in the bulk, and the thermodynamic topology may be reconfigured through defect pair creation/annihilation or the persistence of long-lived defects.

\section{Conclusions}
\label{sec:theend}
In this work, we have established both an analytic and numerical bridge between the microscopic matter content - encoded by the coupling $\alpha$ and the electric charge scale $Q$ - the classical energy conditions, the Duan thermodynamic topology $(\mathcal{F},\varphi^S,\tau)$, and the Ruppeiner thermodynamic geometry. The core analytic results are compact and robust. In particular, we have shown that a small-radius expansion of the Duan vector yields Eq. \eqref{rob}, 
so that the dimensionless combination $\alpha^2|Q|$ naturally classifies the phase structure into three regimes: subcritical ($\alpha^2|Q|<1$), critical ($\alpha^2|Q|=1$) and supercritical ($\alpha^2|Q|>1$). Independently, the leading estimate for the null/weak energy-condition threshold in Eq. \eqref{rnec}
provides a practical diagnostic: if the outer horizon $r_h$ satisfies $r_h>r_{\rm NEC}$, the black hole probes macroscopic EC-violating regions, while $r_h<r_{\rm NEC}$ indicates an EC-satisfying domain. Combining these two diagnostics yields a simple, testable physical picture.

Concretely, in the subcritical regime the SBH branch emerges in a thermodynamically unstable form ($\varphi^S(0^+)>0$, winding $w=-1$) and the phase structure is dominated by first-order SBH-LBH coexistence. In the critical regime ($\alpha^2|Q|=1$), the system is marginal: $\varphi^S(0^+)=0$ allows vertical tangencies in $\tau(r_h)$, with $J_0=\partial_S\varphi^S \to 0$ and $C \to \infty$, thereby realizing genuine second-order critical points at which topological defect pairs can be created or annihilated. In the supercritical regime, the SBH branch emerges locally stable ($\varphi^S(0^+)<0$, winding $w=+1$), the thermodynamic topology becomes richer (with SBH-IBH-LBH sequences appearing generically), and $r_{\rm NEC}$ is small, so horizons typically reside in regions that violate the energy conditions. 

These results carry both conceptual and practical implications. Conceptually, they demonstrate that macroscopic violations of classical energy conditions - governed by the coupling $\alpha$ - are tightly correlated with the system’s ability to realize nontrivial thermodynamic topology and genuine continuous critical phenomena. In particular, the presence of exotic stress-energy components provides the additional degrees of freedom required to generate inflection points and vertical tangencies in $\tau(r_h)$, thereby enabling the emergence of intermediate branches and higher-order critical behavior that cannot arise in standard, energy-condition-preserving settings. We verified these analytic expectations through targeted numerical checks, including parameter scans in $(\alpha,|Q|)$, direct evaluation of the Duan defect map, and computation of $R_{\rm Rup}(r_h)$, obtaining a consistent and coherent picture across representative parameter sets. 

Looking ahead, this framework suggests several promising directions: a systematic exploration of astrophysically relevant parameter ranges could clarify whether Event Horizon Telescope-scale observations are sensitive to thermodynamic or topological signatures, thereby constraining the coupling $\alpha$. On the microscopic side, the positive short-range Ruppeiner curvature points to an interpretation in terms of repulsive microstructure, motivating a statistical treatment that may uncover the effective degrees of freedom governed by the $\alpha$-coupling. More broadly, the analytic classifier developed here can be extended beyond small-radius asymptotics to include rotation, higher-derivative corrections, and alternative matter sectors, providing a unified tool to map the phase structure of black holes in modified gravity. Lastly, the discovery of observable or analogue gravitational signals, such as gravitational wave echoes or modified ringdown spectra, remains an exciting challenge, as does the motion of rotating particles \cite{qq1,qq2,qq3,Turimov:2025qkz,Yunusov:2025chw}. We leave these directions to future work.

\subsection*{Acknowledgements}
The research of GGL is supported by the Ayuda Postdoctoral of the University of Lleida (project code: X25027). GGL gratefully acknowledges the contribution of the LISA Cosmology Working Group (CosWG), as well as support from the COST Actions CA21136 - \textit{Addressing observational tensions in cosmology with systematics and fundamental physics (CosmoVerse)} - CA23130, \textit{Bridging high and low energies in search of quantum gravity (BridgeQG)} and CA21106 -  \textit{COSMIC WISPers in the Dark Universe: Theory, astrophysics and experiments (CosmicWISPers)}.
The authors SKM and MKJ are thankful to the administration of the University of Nizwa for their continuous support and encouragement for this research work.

\bibliography{Ref.bib}

\begin{thebibliography}{76}%
\makeatletter
\providecommand \@ifxundefined [1]{%
 \@ifx{#1\undefined}
}%
\providecommand \@ifnum [1]{%
 \ifnum #1\expandafter \@firstoftwo
 \else \expandafter \@secondoftwo
 \fi
}%
\providecommand \@ifx [1]{%
 \ifx #1\expandafter \@firstoftwo
 \else \expandafter \@secondoftwo
 \fi
}%
\providecommand \natexlab [1]{#1}%
\providecommand \enquote  [1]{``#1''}%
\providecommand \bibnamefont  [1]{#1}%
\providecommand \bibfnamefont [1]{#1}%
\providecommand \citenamefont [1]{#1}%
\providecommand \href@noop [0]{\@secondoftwo}%
\providecommand \href [0]{\begingroup \@sanitize@url \@href}%
\providecommand \@href[1]{\@@startlink{#1}\@@href}%
\providecommand \@@href[1]{\endgroup#1\@@endlink}%
\providecommand \@sanitize@url [0]{\catcode `\\12\catcode `\$12\catcode `\&12\catcode `\#12\catcode `\^12\catcode `\_12\catcode `\%12\relax}%
\providecommand \@@startlink[1]{}%
\providecommand \@@endlink[0]{}%
\providecommand \url  [0]{\begingroup\@sanitize@url \@url }%
\providecommand \@url [1]{\endgroup\@href {#1}{\urlprefix }}%
\providecommand \urlprefix  [0]{URL }%
\providecommand \Eprint [0]{\href }%
\providecommand \doibase [0]{http://dx.doi.org/}%
\providecommand \selectlanguage [0]{\@gobble}%
\providecommand \bibinfo  [0]{\@secondoftwo}%
\providecommand \bibfield  [0]{\@secondoftwo}%
\providecommand \translation [1]{[#1]}%
\providecommand \BibitemOpen [0]{}%
\providecommand \bibitemStop [0]{}%
\providecommand \bibitemNoStop [0]{.\EOS\space}%
\providecommand \EOS [0]{\spacefactor3000\relax}%
\providecommand \BibitemShut  [1]{\csname bibitem#1\endcsname}%
\let\auto@bib@innerbib\@empty
\bibitem [{\citenamefont {Bekenstein}(1973)}]{Bekenstein:1973ur}%
  \BibitemOpen
  \bibfield  {author} {\bibinfo {author} {\bibfnamefont {Jacob~D.}\ \bibnamefont {Bekenstein}},\ }\bibfield  {title} {\enquote {\bibinfo {title} {{Black holes and entropy}},}\ }\href {\doibase 10.1103/PhysRevD.7.2333} {\bibfield  {journal} {\bibinfo  {journal} {Phys. Rev. D}\ }\textbf {\bibinfo {volume} {7}},\ \bibinfo {pages} {2333--2346} (\bibinfo {year} {1973})}\BibitemShut {NoStop}%
\bibitem [{\citenamefont {Hawking}(1975)}]{Hawking:1975vcx}%
  \BibitemOpen
  \bibfield  {author} {\bibinfo {author} {\bibfnamefont {S.~W.}\ \bibnamefont {Hawking}},\ }\bibfield  {title} {\enquote {\bibinfo {title} {{Particle Creation by Black Holes}},}\ }\href {\doibase 10.1007/BF02345020} {\bibfield  {journal} {\bibinfo  {journal} {Commun. Math. Phys.}\ }\textbf {\bibinfo {volume} {43}},\ \bibinfo {pages} {199--220} (\bibinfo {year} {1975})},\ \bibinfo {note} {[Erratum: Commun.Math.Phys. 46, 206 (1976)]}\BibitemShut {NoStop}%
\bibitem [{\citenamefont {Kubiznak}\ and\ \citenamefont {Mann}(2012)}]{Kubiznak:2012wp}%
  \BibitemOpen
  \bibfield  {author} {\bibinfo {author} {\bibfnamefont {David}\ \bibnamefont {Kubiznak}}\ and\ \bibinfo {author} {\bibfnamefont {Robert~B.}\ \bibnamefont {Mann}},\ }\bibfield  {title} {\enquote {\bibinfo {title} {{P-V criticality of charged AdS black holes}},}\ }\href {\doibase 10.1007/JHEP07(2012)033} {\bibfield  {journal} {\bibinfo  {journal} {JHEP}\ }\textbf {\bibinfo {volume} {07}},\ \bibinfo {pages} {033} (\bibinfo {year} {2012})},\ \Eprint {http://arxiv.org/abs/1205.0559} {arXiv:1205.0559 [hep-th]} \BibitemShut {NoStop}%
\bibitem [{\citenamefont {Rani}\ \emph {et~al.}(2023)\citenamefont {Rani}, \citenamefont {Jawad},\ and\ \citenamefont {Hussain}}]{Rani:2023jai}%
  \BibitemOpen
  \bibfield  {author} {\bibinfo {author} {\bibfnamefont {Shamaila}\ \bibnamefont {Rani}}, \bibinfo {author} {\bibfnamefont {Abdul}\ \bibnamefont {Jawad}}, \ and\ \bibinfo {author} {\bibfnamefont {Mazhar}\ \bibnamefont {Hussain}},\ }\bibfield  {title} {\enquote {\bibinfo {title} {{Impact of barrow entropy on geometrothermodynamics of specific black holes}},}\ }\href {\doibase 10.1140/epjc/s10052-023-11857-5} {\bibfield  {journal} {\bibinfo  {journal} {Eur. Phys. J. C}\ }\textbf {\bibinfo {volume} {83}},\ \bibinfo {pages} {710} (\bibinfo {year} {2023})}\BibitemShut {NoStop}%
\bibitem [{\citenamefont {Rani}\ \emph {et~al.}(2022)\citenamefont {Rani}, \citenamefont {Jawad}, \citenamefont {Moradpour},\ and\ \citenamefont {Tanveer}}]{Rani:2022xza}%
  \BibitemOpen
  \bibfield  {author} {\bibinfo {author} {\bibfnamefont {Shamaila}\ \bibnamefont {Rani}}, \bibinfo {author} {\bibfnamefont {Abdul}\ \bibnamefont {Jawad}}, \bibinfo {author} {\bibfnamefont {Hooman}\ \bibnamefont {Moradpour}}, \ and\ \bibinfo {author} {\bibfnamefont {Aqsa}\ \bibnamefont {Tanveer}},\ }\bibfield  {title} {\enquote {\bibinfo {title} {{Tsallis entropy inspires geometric thermodynamics of specific black hole}},}\ }\href {\doibase 10.1140/epjc/s10052-022-10655-9} {\bibfield  {journal} {\bibinfo  {journal} {Eur. Phys. J. C}\ }\textbf {\bibinfo {volume} {82}},\ \bibinfo {pages} {713} (\bibinfo {year} {2022})}\BibitemShut {NoStop}%
\bibitem [{\citenamefont {Luciano}\ and\ \citenamefont {Sheykhi}(2023)}]{Luciano:2023fyr}%
  \BibitemOpen
  \bibfield  {author} {\bibinfo {author} {\bibfnamefont {G.~G.}\ \bibnamefont {Luciano}}\ and\ \bibinfo {author} {\bibfnamefont {A.}~\bibnamefont {Sheykhi}},\ }\bibfield  {title} {\enquote {\bibinfo {title} {Black hole geometrothermodynamics and critical phenomena: A look from tsallis entropy-based perspective},}\ }\href {\doibase 10.1016/j.dark.2023.101319} {\bibfield  {journal} {\bibinfo  {journal} {Phys. Dark Univ.}\ }\textbf {\bibinfo {volume} {42}},\ \bibinfo {pages} {101319} (\bibinfo {year} {2023})}\BibitemShut {NoStop}%
\bibitem [{\citenamefont {{Turimov}}\ \emph {et~al.}(2025)\citenamefont {{Turimov}}, \citenamefont {{Usanov}},\ and\ \citenamefont {{Khamroev}}}]{2025PDU....4801876T}%
  \BibitemOpen
  \bibfield  {author} {\bibinfo {author} {\bibfnamefont {Bobur}\ \bibnamefont {{Turimov}}}, \bibinfo {author} {\bibfnamefont {Sulton}\ \bibnamefont {{Usanov}}}, \ and\ \bibinfo {author} {\bibfnamefont {Yokubjon}\ \bibnamefont {{Khamroev}}},\ }\bibfield  {title} {\enquote {\bibinfo {title} {{Particles acceleration by Bocharova{\textendash}Bronnikov{\textendash}Melnikov{\textendash}Bekenstein black hole}},}\ }\href {\doibase 10.1016/j.dark.2025.101876} {\bibfield  {journal} {\bibinfo  {journal} {Physics of the Dark Universe}\ }\textbf {\bibinfo {volume} {48}},\ \bibinfo {eid} {101876} (\bibinfo {year} {2025})},\ \Eprint {http://arxiv.org/abs/2502.11185} {arXiv:2502.11185 [gr-qc]} \BibitemShut {NoStop}%
\bibitem [{\citenamefont {Luciano}\ and\ \citenamefont {Saridakis}(2023)}]{Luciano:2023bai}%
  \BibitemOpen
  \bibfield  {author} {\bibinfo {author} {\bibfnamefont {G.~G.}\ \bibnamefont {Luciano}}\ and\ \bibinfo {author} {\bibfnamefont {E.}~\bibnamefont {Saridakis}},\ }\bibfield  {title} {\enquote {\bibinfo {title} {P--v criticalities, phase transitions and geometrothermodynamics of charged ads black holes from kaniadakis statistics},}\ }\href {\doibase 10.1007/JHEP12(2023)114} {\bibfield  {journal} {\bibinfo  {journal} {JHEP}\ }\textbf {\bibinfo {volume} {12}},\ \bibinfo {pages} {114} (\bibinfo {year} {2023})}\BibitemShut {NoStop}%
\bibitem [{\citenamefont {Nakarachinda}\ \emph {et~al.}(2025)\citenamefont {Nakarachinda}, \citenamefont {Promsiri}, \citenamefont {Tannukij},\ and\ \citenamefont {Wongjun}}]{Nakarachinda:2022gsb}%
  \BibitemOpen
  \bibfield  {author} {\bibinfo {author} {\bibfnamefont {Ratchaphat}\ \bibnamefont {Nakarachinda}}, \bibinfo {author} {\bibfnamefont {Chatchai}\ \bibnamefont {Promsiri}}, \bibinfo {author} {\bibfnamefont {Lunchakorn}\ \bibnamefont {Tannukij}}, \ and\ \bibinfo {author} {\bibfnamefont {Pitayuth}\ \bibnamefont {Wongjun}},\ }\bibfield  {title} {\enquote {\bibinfo {title} {{Thermodynamics of black holes with R{\'e}nyi entropy from classical gravity}},}\ }\href {\doibase 10.1016/j.nuclphysb.2025.116796} {\bibfield  {journal} {\bibinfo  {journal} {Nucl. Phys. B}\ }\textbf {\bibinfo {volume} {1011}},\ \bibinfo {pages} {116796} (\bibinfo {year} {2025})},\ \Eprint {http://arxiv.org/abs/2211.05989} {arXiv:2211.05989 [gr-qc]} \BibitemShut {NoStop}%
\bibitem [{\citenamefont {Gangopadhyay}\ \emph {et~al.}(2014)\citenamefont {Gangopadhyay}, \citenamefont {Dutta},\ and\ \citenamefont {Saha}}]{Gangopadhyay:2013ofa}%
  \BibitemOpen
  \bibfield  {author} {\bibinfo {author} {\bibfnamefont {Sunandan}\ \bibnamefont {Gangopadhyay}}, \bibinfo {author} {\bibfnamefont {Abhijit}\ \bibnamefont {Dutta}}, \ and\ \bibinfo {author} {\bibfnamefont {Anirban}\ \bibnamefont {Saha}},\ }\bibfield  {title} {\enquote {\bibinfo {title} {{Generalized uncertainty principle and black hole thermodynamics}},}\ }\href {\doibase 10.1007/s10714-013-1661-3} {\bibfield  {journal} {\bibinfo  {journal} {Gen. Rel. Grav.}\ }\textbf {\bibinfo {volume} {46}},\ \bibinfo {pages} {1661} (\bibinfo {year} {2014})},\ \Eprint {http://arxiv.org/abs/1307.7045} {arXiv:1307.7045 [gr-qc]} \BibitemShut {NoStop}%
\bibitem [{\citenamefont {Hassanabadi}\ \emph {et~al.}(2021)\citenamefont {Hassanabadi}, \citenamefont {K{\v{r}}{\'\i}{\v{z}}}, \citenamefont {Chung}, \citenamefont {L{\"u}tf{\"u}o{\u{g}}lu}, \citenamefont {Maghsoodi},\ and\ \citenamefont {Hassanabadi}}]{Hassanabadi:2021kuc}%
  \BibitemOpen
  \bibfield  {author} {\bibinfo {author} {\bibfnamefont {S.}~\bibnamefont {Hassanabadi}}, \bibinfo {author} {\bibfnamefont {J.}~\bibnamefont {K{\v{r}}{\'\i}{\v{z}}}}, \bibinfo {author} {\bibfnamefont {W.~S.}\ \bibnamefont {Chung}}, \bibinfo {author} {\bibfnamefont {B.~C.}\ \bibnamefont {L{\"u}tf{\"u}o{\u{g}}lu}}, \bibinfo {author} {\bibfnamefont {E.}~\bibnamefont {Maghsoodi}}, \ and\ \bibinfo {author} {\bibfnamefont {H.}~\bibnamefont {Hassanabadi}},\ }\bibfield  {title} {\enquote {\bibinfo {title} {{Thermodynamics of the Schwarzschild and Reissner{\textendash}Nordstr{\"o}m black holes under higher-order generalized uncertainty principle}},}\ }\href {\doibase 10.1140/epjp/s13360-021-01933-8} {\bibfield  {journal} {\bibinfo  {journal} {Eur. Phys. J. Plus}\ }\textbf {\bibinfo {volume} {136}},\ \bibinfo {pages} {918} (\bibinfo {year} {2021})},\ \Eprint {http://arxiv.org/abs/2110.01363} {arXiv:2110.01363 [gr-qc]} \BibitemShut {NoStop}%
\bibitem [{\citenamefont {Kim}\ \emph {et~al.}(2008)\citenamefont {Kim}, \citenamefont {Son},\ and\ \citenamefont {Yoon}}]{Kim:2007hf}%
  \BibitemOpen
  \bibfield  {author} {\bibinfo {author} {\bibfnamefont {Wontae}\ \bibnamefont {Kim}}, \bibinfo {author} {\bibfnamefont {Edwin~J.}\ \bibnamefont {Son}}, \ and\ \bibinfo {author} {\bibfnamefont {Myungseok}\ \bibnamefont {Yoon}},\ }\bibfield  {title} {\enquote {\bibinfo {title} {{Thermodynamics of a black hole based on a generalized uncertainty principle}},}\ }\href {\doibase 10.1088/1126-6708/2008/01/035} {\bibfield  {journal} {\bibinfo  {journal} {JHEP}\ }\textbf {\bibinfo {volume} {01}},\ \bibinfo {pages} {035} (\bibinfo {year} {2008})},\ \Eprint {http://arxiv.org/abs/0711.0786} {arXiv:0711.0786 [gr-qc]} \BibitemShut {NoStop}%
\bibitem [{\citenamefont {Sekhmani}\ \emph {et~al.}(2025)\citenamefont {Sekhmani}, \citenamefont {Luciano}, \citenamefont {Gashti},\ and\ \citenamefont {Baruah}}]{Sekhmani:2025zwc}%
  \BibitemOpen
  \bibfield  {author} {\bibinfo {author} {\bibfnamefont {Y.}~\bibnamefont {Sekhmani}}, \bibinfo {author} {\bibfnamefont {G.~G.}\ \bibnamefont {Luciano}}, \bibinfo {author} {\bibfnamefont {S.~N.}\ \bibnamefont {Gashti}}, \ and\ \bibinfo {author} {\bibfnamefont {A.}~\bibnamefont {Baruah}},\ }\bibfield  {title} {\enquote {\bibinfo {title} {{Infrared Extended Uncertainty Principle corrections and quintessence-induced topology of Reissner{\textendash}Nordstr{\"o}m AdS black holes}},}\ }\href {\doibase 10.1016/j.aop.2025.170187} {\bibfield  {journal} {\bibinfo  {journal} {Annals Phys.}\ }\textbf {\bibinfo {volume} {481}},\ \bibinfo {pages} {170187} (\bibinfo {year} {2025})},\ \Eprint {http://arxiv.org/abs/2508.14953} {arXiv:2508.14953 [gr-qc]} \BibitemShut {NoStop}%
\bibitem [{\citenamefont {Chamblin}\ \emph {et~al.}(1999)\citenamefont {Chamblin}, \citenamefont {Emparan}, \citenamefont {Johnson},\ and\ \citenamefont {Myers}}]{Chamblin:1999hg}%
  \BibitemOpen
  \bibfield  {author} {\bibinfo {author} {\bibfnamefont {Andrew}\ \bibnamefont {Chamblin}}, \bibinfo {author} {\bibfnamefont {Roberto}\ \bibnamefont {Emparan}}, \bibinfo {author} {\bibfnamefont {Clifford~V.}\ \bibnamefont {Johnson}}, \ and\ \bibinfo {author} {\bibfnamefont {Robert~C.}\ \bibnamefont {Myers}},\ }\bibfield  {title} {\enquote {\bibinfo {title} {{Holography, thermodynamics and fluctuations of charged AdS black holes}},}\ }\href {\doibase 10.1103/PhysRevD.60.104026} {\bibfield  {journal} {\bibinfo  {journal} {Phys. Rev. D}\ }\textbf {\bibinfo {volume} {60}},\ \bibinfo {pages} {104026} (\bibinfo {year} {1999})},\ \Eprint {http://arxiv.org/abs/hep-th/9904197} {arXiv:hep-th/9904197} \BibitemShut {NoStop}%
\bibitem [{\citenamefont {Caldarelli}\ \emph {et~al.}(2000)\citenamefont {Caldarelli}, \citenamefont {Cognola},\ and\ \citenamefont {Klemm}}]{Caldarelli:1999xj}%
  \BibitemOpen
  \bibfield  {author} {\bibinfo {author} {\bibfnamefont {Marco~M.}\ \bibnamefont {Caldarelli}}, \bibinfo {author} {\bibfnamefont {Guido}\ \bibnamefont {Cognola}}, \ and\ \bibinfo {author} {\bibfnamefont {Dietmar}\ \bibnamefont {Klemm}},\ }\bibfield  {title} {\enquote {\bibinfo {title} {{Thermodynamics of Kerr-Newman-AdS black holes and conformal field theories}},}\ }\href {\doibase 10.1088/0264-9381/17/2/310} {\bibfield  {journal} {\bibinfo  {journal} {Class. Quant. Grav.}\ }\textbf {\bibinfo {volume} {17}},\ \bibinfo {pages} {399--420} (\bibinfo {year} {2000})},\ \Eprint {http://arxiv.org/abs/hep-th/9908022} {arXiv:hep-th/9908022} \BibitemShut {NoStop}%
\bibitem [{\citenamefont {Padmanabhan}(2010)}]{Padmanabhan:2009vy}%
  \BibitemOpen
  \bibfield  {author} {\bibinfo {author} {\bibfnamefont {T.}~\bibnamefont {Padmanabhan}},\ }\bibfield  {title} {\enquote {\bibinfo {title} {{Thermodynamical Aspects of Gravity: New insights}},}\ }\href {\doibase 10.1088/0034-4885/73/4/046901} {\bibfield  {journal} {\bibinfo  {journal} {Rept. Prog. Phys.}\ }\textbf {\bibinfo {volume} {73}},\ \bibinfo {pages} {046901} (\bibinfo {year} {2010})},\ \Eprint {http://arxiv.org/abs/0911.5004} {arXiv:0911.5004 [gr-qc]} \BibitemShut {NoStop}%
\bibitem [{\citenamefont {Cai}\ and\ \citenamefont {Kim}(2005)}]{Cai:2005ra}%
  \BibitemOpen
  \bibfield  {author} {\bibinfo {author} {\bibfnamefont {Rong-Gen}\ \bibnamefont {Cai}}\ and\ \bibinfo {author} {\bibfnamefont {Sang~Pyo}\ \bibnamefont {Kim}},\ }\bibfield  {title} {\enquote {\bibinfo {title} {{First law of thermodynamics and Friedmann equations of Friedmann-Robertson-Walker universe}},}\ }\href {\doibase 10.1088/1126-6708/2005/02/050} {\bibfield  {journal} {\bibinfo  {journal} {JHEP}\ }\textbf {\bibinfo {volume} {02}},\ \bibinfo {pages} {050} (\bibinfo {year} {2005})},\ \Eprint {http://arxiv.org/abs/hep-th/0501055} {arXiv:hep-th/0501055} \BibitemShut {NoStop}%
\bibitem [{\citenamefont {Wald}(2001)}]{Wald:1999vt}%
  \BibitemOpen
  \bibfield  {author} {\bibinfo {author} {\bibfnamefont {Robert~M.}\ \bibnamefont {Wald}},\ }\bibfield  {title} {\enquote {\bibinfo {title} {{The thermodynamics of black holes}},}\ }\href {\doibase 10.12942/lrr-2001-6} {\bibfield  {journal} {\bibinfo  {journal} {Living Rev. Rel.}\ }\textbf {\bibinfo {volume} {4}},\ \bibinfo {pages} {6} (\bibinfo {year} {2001})},\ \Eprint {http://arxiv.org/abs/gr-qc/9912119} {arXiv:gr-qc/9912119} \BibitemShut {NoStop}%
\bibitem [{\citenamefont {Kubiznak}\ \emph {et~al.}(2017)\citenamefont {Kubiznak}, \citenamefont {Mann},\ and\ \citenamefont {Teo}}]{Kubiznak:2016qmn}%
  \BibitemOpen
  \bibfield  {author} {\bibinfo {author} {\bibfnamefont {David}\ \bibnamefont {Kubiznak}}, \bibinfo {author} {\bibfnamefont {Robert~B.}\ \bibnamefont {Mann}}, \ and\ \bibinfo {author} {\bibfnamefont {Mae}\ \bibnamefont {Teo}},\ }\bibfield  {title} {\enquote {\bibinfo {title} {{Black hole chemistry: thermodynamics with Lambda}},}\ }\href {\doibase 10.1088/1361-6382/aa5c69} {\bibfield  {journal} {\bibinfo  {journal} {Class. Quant. Grav.}\ }\textbf {\bibinfo {volume} {34}},\ \bibinfo {pages} {063001} (\bibinfo {year} {2017})},\ \Eprint {http://arxiv.org/abs/1608.06147} {arXiv:1608.06147 [hep-th]} \BibitemShut {NoStop}%
\bibitem [{\citenamefont {Dolan}(2011)}]{Dolan:2011xt}%
  \BibitemOpen
  \bibfield  {author} {\bibinfo {author} {\bibfnamefont {Brian~P.}\ \bibnamefont {Dolan}},\ }\bibfield  {title} {\enquote {\bibinfo {title} {{Pressure and volume in the first law of black hole thermodynamics}},}\ }\href {\doibase 10.1088/0264-9381/28/23/235017} {\bibfield  {journal} {\bibinfo  {journal} {Class. Quant. Grav.}\ }\textbf {\bibinfo {volume} {28}},\ \bibinfo {pages} {235017} (\bibinfo {year} {2011})},\ \Eprint {http://arxiv.org/abs/1106.6260} {arXiv:1106.6260 [gr-qc]} \BibitemShut {NoStop}%
\bibitem [{\citenamefont {Cvetic}\ \emph {et~al.}(2011)\citenamefont {Cvetic}, \citenamefont {Gibbons}, \citenamefont {Kubiznak},\ and\ \citenamefont {Pope}}]{Cvetic:2010jb}%
  \BibitemOpen
  \bibfield  {author} {\bibinfo {author} {\bibfnamefont {M.}~\bibnamefont {Cvetic}}, \bibinfo {author} {\bibfnamefont {G.~W.}\ \bibnamefont {Gibbons}}, \bibinfo {author} {\bibfnamefont {D.}~\bibnamefont {Kubiznak}}, \ and\ \bibinfo {author} {\bibfnamefont {C.~N.}\ \bibnamefont {Pope}},\ }\bibfield  {title} {\enquote {\bibinfo {title} {{Black Hole Enthalpy and an Entropy Inequality for the Thermodynamic Volume}},}\ }\href {\doibase 10.1103/PhysRevD.84.024037} {\bibfield  {journal} {\bibinfo  {journal} {Phys. Rev. D}\ }\textbf {\bibinfo {volume} {84}},\ \bibinfo {pages} {024037} (\bibinfo {year} {2011})},\ \Eprint {http://arxiv.org/abs/1012.2888} {arXiv:1012.2888 [hep-th]} \BibitemShut {NoStop}%
\bibitem [{\citenamefont {Gunasekaran}\ \emph {et~al.}(2012)\citenamefont {Gunasekaran}, \citenamefont {Mann},\ and\ \citenamefont {Kubiznak}}]{Gunasekaran:2012dq}%
  \BibitemOpen
  \bibfield  {author} {\bibinfo {author} {\bibfnamefont {Sharmila}\ \bibnamefont {Gunasekaran}}, \bibinfo {author} {\bibfnamefont {Robert~B.}\ \bibnamefont {Mann}}, \ and\ \bibinfo {author} {\bibfnamefont {David}\ \bibnamefont {Kubiznak}},\ }\bibfield  {title} {\enquote {\bibinfo {title} {{Extended phase space thermodynamics for charged and rotating black holes and Born-Infeld vacuum polarization}},}\ }\href {\doibase 10.1007/JHEP11(2012)110} {\bibfield  {journal} {\bibinfo  {journal} {JHEP}\ }\textbf {\bibinfo {volume} {11}},\ \bibinfo {pages} {110} (\bibinfo {year} {2012})},\ \Eprint {http://arxiv.org/abs/1208.6251} {arXiv:1208.6251 [hep-th]} \BibitemShut {NoStop}%
\bibitem [{\citenamefont {Hayward}(1998)}]{Hayward:1997jp}%
  \BibitemOpen
  \bibfield  {author} {\bibinfo {author} {\bibfnamefont {Sean~A.}\ \bibnamefont {Hayward}},\ }\bibfield  {title} {\enquote {\bibinfo {title} {{Unified first law of black hole dynamics and relativistic thermodynamics}},}\ }\href {\doibase 10.1088/0264-9381/15/10/017} {\bibfield  {journal} {\bibinfo  {journal} {Class. Quant. Grav.}\ }\textbf {\bibinfo {volume} {15}},\ \bibinfo {pages} {3147--3162} (\bibinfo {year} {1998})},\ \Eprint {http://arxiv.org/abs/gr-qc/9710089} {arXiv:gr-qc/9710089} \BibitemShut {NoStop}%
\bibitem [{\citenamefont {Capozziello}\ \emph {et~al.}(2023)\citenamefont {Capozziello}, \citenamefont {D'Agostino}, \citenamefont {Lapponi},\ and\ \citenamefont {Luongo}}]{Capozziello:2022ygp}%
  \BibitemOpen
  \bibfield  {author} {\bibinfo {author} {\bibfnamefont {Salvatore}\ \bibnamefont {Capozziello}}, \bibinfo {author} {\bibfnamefont {Rocco}\ \bibnamefont {D'Agostino}}, \bibinfo {author} {\bibfnamefont {Alessio}\ \bibnamefont {Lapponi}}, \ and\ \bibinfo {author} {\bibfnamefont {Orlando}\ \bibnamefont {Luongo}},\ }\bibfield  {title} {\enquote {\bibinfo {title} {{Black hole thermodynamics from logotropic fluids}},}\ }\href {\doibase 10.1140/epjc/s10052-023-11319-y} {\bibfield  {journal} {\bibinfo  {journal} {Eur. Phys. J. C}\ }\textbf {\bibinfo {volume} {83}},\ \bibinfo {pages} {175} (\bibinfo {year} {2023})},\ \Eprint {http://arxiv.org/abs/2210.02431} {arXiv:2210.02431 [gr-qc]} \BibitemShut {NoStop}%
\bibitem [{\citenamefont {Ghaffari}\ and\ \citenamefont {Luciano}(2025)}]{Ghaffari:2025qmv}%
  \BibitemOpen
  \bibfield  {author} {\bibinfo {author} {\bibfnamefont {S.}~\bibnamefont {Ghaffari}}\ and\ \bibinfo {author} {\bibfnamefont {G.~G.}\ \bibnamefont {Luciano}},\ }\bibfield  {title} {\enquote {\bibinfo {title} {{Black hole thermodynamics in Harada{\textquoteright}s inspired theory of gravity: stability, phase structure and geometrothermodynamics}},}\ }\href {\doibase 10.1140/epjc/s10052-025-14515-0} {\bibfield  {journal} {\bibinfo  {journal} {Eur. Phys. J. C}\ }\textbf {\bibinfo {volume} {85}},\ \bibinfo {pages} {785} (\bibinfo {year} {2025})},\ \Eprint {http://arxiv.org/abs/2505.06560} {arXiv:2505.06560 [gr-qc]} \BibitemShut {NoStop}%
\bibitem [{\citenamefont {Duan}\ \emph {et~al.}(1998)\citenamefont {Duan}, \citenamefont {Jiang},\ and\ \citenamefont {Yang}}]{Duan:1998it}%
  \BibitemOpen
  \bibfield  {author} {\bibinfo {author} {\bibfnamefont {Yi-shi}\ \bibnamefont {Duan}}, \bibinfo {author} {\bibfnamefont {Ying}\ \bibnamefont {Jiang}}, \ and\ \bibinfo {author} {\bibfnamefont {Guo-hong}\ \bibnamefont {Yang}},\ }\bibfield  {title} {\enquote {\bibinfo {title} {{The Topological quantization and the branch process of the (k-1)-dimensional topological defects}},}\ }\href {\doibase https://doi.org/10.48550/arXiv.hep-th/9810111} {\bibfield  {journal} {\bibinfo  {journal} {arXiv}\ } (\bibinfo {year} {1998}),\ https://doi.org/10.48550/arXiv.hep-th/9810111},\ \Eprint {http://arxiv.org/abs/hep-th/9810111} {hep-th/9810111} \BibitemShut {NoStop}%
\bibitem [{\citenamefont {Wei}\ and\ \citenamefont {Liu}(2022{\natexlab{a}})}]{Wei:2022dzw}%
  \BibitemOpen
  \bibfield  {author} {\bibinfo {author} {\bibfnamefont {Shao-Wen}\ \bibnamefont {Wei}}\ and\ \bibinfo {author} {\bibfnamefont {Yu-Xiao}\ \bibnamefont {Liu}},\ }\bibfield  {title} {\enquote {\bibinfo {title} {{Topology of black hole thermodynamics}},}\ }\href {\doibase 10.1103/PhysRevD.105.104003} {\bibfield  {journal} {\bibinfo  {journal} {Phys. Rev. D}\ }\textbf {\bibinfo {volume} {105}},\ \bibinfo {pages} {104003} (\bibinfo {year} {2022}{\natexlab{a}})},\ \Eprint {http://arxiv.org/abs/2112.01706} {arXiv:2112.01706 [gr-qc]} \BibitemShut {NoStop}%
\bibitem [{\citenamefont {Wei}\ \emph {et~al.}(2022{\natexlab{a}})\citenamefont {Wei}, \citenamefont {Liu},\ and\ \citenamefont {Mann}}]{Wei:2022dzw2}%
  \BibitemOpen
  \bibfield  {author} {\bibinfo {author} {\bibfnamefont {Shao-Wen}\ \bibnamefont {Wei}}, \bibinfo {author} {\bibfnamefont {Yu-Xiao}\ \bibnamefont {Liu}}, \ and\ \bibinfo {author} {\bibfnamefont {Robert~B.}\ \bibnamefont {Mann}},\ }\bibfield  {title} {\enquote {\bibinfo {title} {{Black Hole Solutions as Topological Thermodynamic Defects}},}\ }\href {\doibase 10.1103/PhysRevLett.129.191101} {\bibfield  {journal} {\bibinfo  {journal} {Phys. Rev. Lett.}\ }\textbf {\bibinfo {volume} {129}},\ \bibinfo {pages} {191101} (\bibinfo {year} {2022}{\natexlab{a}})},\ \Eprint {http://arxiv.org/abs/2208.01932} {arXiv:2208.01932 [gr-qc]} \BibitemShut {NoStop}%
\bibitem [{\citenamefont {Noori~Gashti}\ \emph {et~al.}(2025)\citenamefont {Noori~Gashti}, \citenamefont {Sakall{\i}},\ and\ \citenamefont {Pourhassan}}]{NooriGashti:2024gnc}%
  \BibitemOpen
  \bibfield  {author} {\bibinfo {author} {\bibfnamefont {Saeed}\ \bibnamefont {Noori~Gashti}}, \bibinfo {author} {\bibfnamefont {{\.I}zzet}\ \bibnamefont {Sakall{\i}}}, \ and\ \bibinfo {author} {\bibfnamefont {Behnam}\ \bibnamefont {Pourhassan}},\ }\bibfield  {title} {\enquote {\bibinfo {title} {{Thermodynamic topology, photon spheres, and evidence for weak gravity conjecture in charged black holes with perfect fluid within Rastall theory}},}\ }\href {\doibase 10.1016/j.physletb.2025.139862} {\bibfield  {journal} {\bibinfo  {journal} {Phys. Lett. B}\ }\textbf {\bibinfo {volume} {869}},\ \bibinfo {pages} {139862} (\bibinfo {year} {2025})},\ \Eprint {http://arxiv.org/abs/2410.14492} {arXiv:2410.14492 [hep-th]} \BibitemShut {NoStop}%
\bibitem [{\citenamefont {Anand}\ and\ \citenamefont {Noori~Gashti}(2025{\natexlab{a}})}]{Anand:2025ttx}%
  \BibitemOpen
  \bibfield  {author} {\bibinfo {author} {\bibfnamefont {Ankit}\ \bibnamefont {Anand}}\ and\ \bibinfo {author} {\bibfnamefont {Saeed}\ \bibnamefont {Noori~Gashti}},\ }\bibfield  {title} {\enquote {\bibinfo {title} {{Van der Waals black holes: Universality, quantum corrections, and topological classifications}},}\ }\href {\doibase 10.1016/j.dark.2025.102018} {\bibfield  {journal} {\bibinfo  {journal} {Phys. Dark Univ.}\ }\textbf {\bibinfo {volume} {49}},\ \bibinfo {pages} {102018} (\bibinfo {year} {2025}{\natexlab{a}})},\ \Eprint {http://arxiv.org/abs/2507.21663} {arXiv:2507.21663 [gr-qc]} \BibitemShut {NoStop}%
\bibitem [{\citenamefont {Anand}\ \emph {et~al.}(2025)\citenamefont {Anand}, \citenamefont {Noori~Gashti},\ and\ \citenamefont {Singh}}]{Anand:2025mlc}%
  \BibitemOpen
  \bibfield  {author} {\bibinfo {author} {\bibfnamefont {Ankit}\ \bibnamefont {Anand}}, \bibinfo {author} {\bibfnamefont {Saeed}\ \bibnamefont {Noori~Gashti}}, \ and\ \bibinfo {author} {\bibfnamefont {Aditya}\ \bibnamefont {Singh}},\ }\bibfield  {title} {\enquote {\bibinfo {title} {{Thermodynamic curvature and topological insights of Hayward black holes with string fluids}},}\ }\href {\doibase 10.1016/j.dark.2025.101994} {\bibfield  {journal} {\bibinfo  {journal} {Phys. Dark Univ.}\ }\textbf {\bibinfo {volume} {49}},\ \bibinfo {pages} {101994} (\bibinfo {year} {2025})},\ \Eprint {http://arxiv.org/abs/2506.23736} {arXiv:2506.23736 [gr-qc]} \BibitemShut {NoStop}%
\bibitem [{\citenamefont {Afshar}\ \emph {et~al.}(2025)\citenamefont {Afshar}, \citenamefont {Alipour}, \citenamefont {Noori~Gashti},\ and\ \citenamefont {Sadeghi}}]{Afshar:2025oqn}%
  \BibitemOpen
  \bibfield  {author} {\bibinfo {author} {\bibfnamefont {Mohammad Ali~S.}\ \bibnamefont {Afshar}}, \bibinfo {author} {\bibfnamefont {Mohammad~Reza}\ \bibnamefont {Alipour}}, \bibinfo {author} {\bibfnamefont {Saeed}\ \bibnamefont {Noori~Gashti}}, \ and\ \bibinfo {author} {\bibfnamefont {Jafar}\ \bibnamefont {Sadeghi}},\ }\bibfield  {title} {\enquote {\bibinfo {title} {{A Deep Dive into classical and Topological CFT Thermodynamics in Lifshitz and Hyperscaling Violating Black Holes}},}\ }\href {\doibase 10.1002/prop.70030} {\bibfield  {journal} {\bibinfo  {journal} {Fortsch. Phys.}\ }\textbf {\bibinfo {volume} {2025}},\ \bibinfo {pages} {e70030} (\bibinfo {year} {2025})},\ \Eprint {http://arxiv.org/abs/2505.15314} {arXiv:2505.15314 [hep-th]} \BibitemShut {NoStop}%
\bibitem [{\citenamefont {Anand}\ and\ \citenamefont {Noori~Gashti}(2025{\natexlab{b}})}]{Anand:2025qow}%
  \BibitemOpen
  \bibfield  {author} {\bibinfo {author} {\bibfnamefont {Ankit}\ \bibnamefont {Anand}}\ and\ \bibinfo {author} {\bibfnamefont {Saeed}\ \bibnamefont {Noori~Gashti}},\ }\bibfield  {title} {\enquote {\bibinfo {title} {{Universality relation and thermodynamic topology with three-parameter entropy model}},}\ }\href {\doibase 10.1016/j.dark.2025.101916} {\bibfield  {journal} {\bibinfo  {journal} {Phys. Dark Univ.}\ }\textbf {\bibinfo {volume} {48}},\ \bibinfo {pages} {101916} (\bibinfo {year} {2025}{\natexlab{b}})}\BibitemShut {NoStop}%
\bibitem [{\citenamefont {Weinhold}(1975)}]{Wein1}%
  \BibitemOpen
  \bibfield  {author} {\bibinfo {author} {\bibfnamefont {F.}~\bibnamefont {Weinhold}},\ }\bibfield  {title} {\enquote {\bibinfo {title} {Metric geometry of equilibrium thermodynamics},}\ }\href {\doibase 10.1063/1.431689} {\bibfield  {journal} {\bibinfo  {journal} {J. Chem. Phys.}\ }\textbf {\bibinfo {volume} {63}},\ \bibinfo {pages} {2479} (\bibinfo {year} {1975})}\BibitemShut {NoStop}%
\bibitem [{\citenamefont {Ruppeiner}(1979)}]{Rupp1}%
  \BibitemOpen
  \bibfield  {author} {\bibinfo {author} {\bibfnamefont {G.}~\bibnamefont {Ruppeiner}},\ }\bibfield  {title} {\enquote {\bibinfo {title} {Thermodynamics: A riemannian geometric model},}\ }\href {\doibase 10.1103/PhysRevA.20.1608} {\bibfield  {journal} {\bibinfo  {journal} {Phys. Rev. A}\ }\textbf {\bibinfo {volume} {20}},\ \bibinfo {pages} {1608} (\bibinfo {year} {1979})}\BibitemShut {NoStop}%
\bibitem [{\citenamefont {Ruppeiner}(1995)}]{Rupp2}%
  \BibitemOpen
  \bibfield  {author} {\bibinfo {author} {\bibfnamefont {G.}~\bibnamefont {Ruppeiner}},\ }\bibfield  {title} {\enquote {\bibinfo {title} {Riemannian geometry in thermodynamic fluctuation theory},}\ }\href {\doibase 10.1103/RevModPhys.67.605} {\bibfield  {journal} {\bibinfo  {journal} {Rev. Mod. Phys.}\ }\textbf {\bibinfo {volume} {67}},\ \bibinfo {pages} {605} (\bibinfo {year} {1995})}\BibitemShut {NoStop}%
\bibitem [{\citenamefont {Wei}\ and\ \citenamefont {Liu}(2013)}]{Wei:2012ui}%
  \BibitemOpen
  \bibfield  {author} {\bibinfo {author} {\bibfnamefont {Shao-Wen}\ \bibnamefont {Wei}}\ and\ \bibinfo {author} {\bibfnamefont {Yu-Xiao}\ \bibnamefont {Liu}},\ }\bibfield  {title} {\enquote {\bibinfo {title} {{Critical phenomena and thermodynamic geometry of charged Gauss-Bonnet AdS black holes}},}\ }\href {\doibase 10.1103/PhysRevD.87.044014} {\bibfield  {journal} {\bibinfo  {journal} {Phys. Rev. D}\ }\textbf {\bibinfo {volume} {87}},\ \bibinfo {pages} {044014} (\bibinfo {year} {2013})},\ \Eprint {http://arxiv.org/abs/1209.1707} {arXiv:1209.1707 [gr-qc]} \BibitemShut {NoStop}%
\bibitem [{\citenamefont {Zhang}\ \emph {et~al.}(2015{\natexlab{a}})\citenamefont {Zhang}, \citenamefont {Cai},\ and\ \citenamefont {Yu}}]{Zhang:2014uoa}%
  \BibitemOpen
  \bibfield  {author} {\bibinfo {author} {\bibfnamefont {Jia-Lin}\ \bibnamefont {Zhang}}, \bibinfo {author} {\bibfnamefont {Rong-Gen}\ \bibnamefont {Cai}}, \ and\ \bibinfo {author} {\bibfnamefont {Hongwei}\ \bibnamefont {Yu}},\ }\bibfield  {title} {\enquote {\bibinfo {title} {{Phase transition and thermodynamical geometry for Schwarzschild AdS black hole in AdS$_{5}$ {\texttimes} S$^{5}$ spacetime}},}\ }\href {\doibase 10.1007/JHEP02(2015)143} {\bibfield  {journal} {\bibinfo  {journal} {JHEP}\ }\textbf {\bibinfo {volume} {02}},\ \bibinfo {pages} {143} (\bibinfo {year} {2015}{\natexlab{a}})},\ \Eprint {http://arxiv.org/abs/1409.5305} {arXiv:1409.5305 [hep-th]} \BibitemShut {NoStop}%
\bibitem [{\citenamefont {Hendi}\ \emph {et~al.}(2015{\natexlab{a}})\citenamefont {Hendi}, \citenamefont {Panahiyan},\ and\ \citenamefont {Eslam~Panah}}]{Hendi:2014kha}%
  \BibitemOpen
  \bibfield  {author} {\bibinfo {author} {\bibfnamefont {S.~H.}\ \bibnamefont {Hendi}}, \bibinfo {author} {\bibfnamefont {S.}~\bibnamefont {Panahiyan}}, \ and\ \bibinfo {author} {\bibfnamefont {B.}~\bibnamefont {Eslam~Panah}},\ }\bibfield  {title} {\enquote {\bibinfo {title} {{P{\textendash}V criticality and geometrical thermodynamics of black holes with Born{\textendash}Infeld type nonlinear electrodynamics}},}\ }\href {\doibase 10.1142/S0218271816500103} {\bibfield  {journal} {\bibinfo  {journal} {Int. J. Mod. Phys. D}\ }\textbf {\bibinfo {volume} {25}},\ \bibinfo {pages} {1650010} (\bibinfo {year} {2015}{\natexlab{a}})},\ \Eprint {http://arxiv.org/abs/1410.0352} {arXiv:1410.0352 [gr-qc]} \BibitemShut {NoStop}%
\bibitem [{\citenamefont {Hendi}\ \emph {et~al.}(2015{\natexlab{b}})\citenamefont {Hendi}, \citenamefont {Sheykhi}, \citenamefont {Panahiyan},\ and\ \citenamefont {Eslam~Panah}}]{Hendi:2015xya}%
  \BibitemOpen
  \bibfield  {author} {\bibinfo {author} {\bibfnamefont {S.~H.}\ \bibnamefont {Hendi}}, \bibinfo {author} {\bibfnamefont {A.}~\bibnamefont {Sheykhi}}, \bibinfo {author} {\bibfnamefont {S.}~\bibnamefont {Panahiyan}}, \ and\ \bibinfo {author} {\bibfnamefont {B.}~\bibnamefont {Eslam~Panah}},\ }\bibfield  {title} {\enquote {\bibinfo {title} {{Phase transition and thermodynamic geometry of Einstein-Maxwell-dilaton black holes}},}\ }\href {\doibase 10.1103/PhysRevD.92.064028} {\bibfield  {journal} {\bibinfo  {journal} {Phys. Rev. D}\ }\textbf {\bibinfo {volume} {92}},\ \bibinfo {pages} {064028} (\bibinfo {year} {2015}{\natexlab{b}})},\ \Eprint {http://arxiv.org/abs/1509.08593} {arXiv:1509.08593 [hep-th]} \BibitemShut {NoStop}%
\bibitem [{\citenamefont {Zhang}\ \emph {et~al.}(2015{\natexlab{b}})\citenamefont {Zhang}, \citenamefont {Cai},\ and\ \citenamefont {Yu}}]{Zhang:2015ova}%
  \BibitemOpen
  \bibfield  {author} {\bibinfo {author} {\bibfnamefont {Jia-Lin}\ \bibnamefont {Zhang}}, \bibinfo {author} {\bibfnamefont {Rong-Gen}\ \bibnamefont {Cai}}, \ and\ \bibinfo {author} {\bibfnamefont {Hongwei}\ \bibnamefont {Yu}},\ }\bibfield  {title} {\enquote {\bibinfo {title} {{Phase transition and thermodynamical geometry of Reissner-Nordstr{\"o}m-AdS black holes in extended phase space}},}\ }\href {\doibase 10.1103/PhysRevD.91.044028} {\bibfield  {journal} {\bibinfo  {journal} {Phys. Rev. D}\ }\textbf {\bibinfo {volume} {91}},\ \bibinfo {pages} {044028} (\bibinfo {year} {2015}{\natexlab{b}})},\ \Eprint {http://arxiv.org/abs/1502.01428} {arXiv:1502.01428 [hep-th]} \BibitemShut {NoStop}%
\bibitem [{\citenamefont {Eslam~Panah}(2018)}]{EslamPanah:2018ums}%
  \BibitemOpen
  \bibfield  {author} {\bibinfo {author} {\bibfnamefont {B.}~\bibnamefont {Eslam~Panah}},\ }\bibfield  {title} {\enquote {\bibinfo {title} {{Effects of energy dependent spacetime on geometrical thermodynamics and heat engine of black holes: gravity's rainbow}},}\ }\href {\doibase 10.1016/j.physletb.2018.10.042} {\bibfield  {journal} {\bibinfo  {journal} {Phys. Lett. B}\ }\textbf {\bibinfo {volume} {787}},\ \bibinfo {pages} {45--55} (\bibinfo {year} {2018})},\ \Eprint {http://arxiv.org/abs/1805.03014} {arXiv:1805.03014 [hep-th]} \BibitemShut {NoStop}%
\bibitem [{\citenamefont {Ghosh}\ and\ \citenamefont {Bhamidipati}(2020{\natexlab{a}})}]{Ghosh:2019pwy}%
  \BibitemOpen
  \bibfield  {author} {\bibinfo {author} {\bibfnamefont {Aritra}\ \bibnamefont {Ghosh}}\ and\ \bibinfo {author} {\bibfnamefont {Chandrasekhar}\ \bibnamefont {Bhamidipati}},\ }\bibfield  {title} {\enquote {\bibinfo {title} {{Thermodynamic geometry for charged Gauss-Bonnet black holes in AdS spacetimes}},}\ }\href {\doibase 10.1103/PhysRevD.101.046005} {\bibfield  {journal} {\bibinfo  {journal} {Phys. Rev. D}\ }\textbf {\bibinfo {volume} {101}},\ \bibinfo {pages} {046005} (\bibinfo {year} {2020}{\natexlab{a}})},\ \Eprint {http://arxiv.org/abs/1911.06280} {arXiv:1911.06280 [gr-qc]} \BibitemShut {NoStop}%
\bibitem [{\citenamefont {Hosseini~Mansoori}(2021)}]{HosseiniMansoori:2020yfj}%
  \BibitemOpen
  \bibfield  {author} {\bibinfo {author} {\bibfnamefont {Seyed~Ali}\ \bibnamefont {Hosseini~Mansoori}},\ }\bibfield  {title} {\enquote {\bibinfo {title} {{Thermodynamic geometry of the novel 4-D Gauss{\textendash}Bonnet AdS black hole}},}\ }\href {\doibase 10.1016/j.dark.2021.100776} {\bibfield  {journal} {\bibinfo  {journal} {Phys. Dark Univ.}\ }\textbf {\bibinfo {volume} {31}},\ \bibinfo {pages} {100776} (\bibinfo {year} {2021})},\ \Eprint {http://arxiv.org/abs/2003.13382} {arXiv:2003.13382 [gr-qc]} \BibitemShut {NoStop}%
\bibitem [{\citenamefont {Born}\ and\ \citenamefont {Infeld}(1934)}]{Born:1934gh}%
  \BibitemOpen
  \bibfield  {author} {\bibinfo {author} {\bibfnamefont {M.}~\bibnamefont {Born}}\ and\ \bibinfo {author} {\bibfnamefont {L.}~\bibnamefont {Infeld}},\ }\bibfield  {title} {\enquote {\bibinfo {title} {{Foundations of the new field theory}},}\ }\href {\doibase 10.1098/rspa.1934.0059} {\bibfield  {journal} {\bibinfo  {journal} {Proc. Roy. Soc. Lond. A}\ }\textbf {\bibinfo {volume} {144}},\ \bibinfo {pages} {425--451} (\bibinfo {year} {1934})}\BibitemShut {NoStop}%
\bibitem [{\citenamefont {Fradkin}\ and\ \citenamefont {Tseytlin}(1985)}]{Fradkin:1985qd}%
  \BibitemOpen
  \bibfield  {author} {\bibinfo {author} {\bibfnamefont {E.~S.}\ \bibnamefont {Fradkin}}\ and\ \bibinfo {author} {\bibfnamefont {Arkady~A.}\ \bibnamefont {Tseytlin}},\ }\bibfield  {title} {\enquote {\bibinfo {title} {{Nonlinear Electrodynamics from Quantized Strings}},}\ }\href {\doibase 10.1016/0370-2693(85)90205-9} {\bibfield  {journal} {\bibinfo  {journal} {Phys. Lett. B}\ }\textbf {\bibinfo {volume} {163}},\ \bibinfo {pages} {123--130} (\bibinfo {year} {1985})}\BibitemShut {NoStop}%
\bibitem [{\citenamefont {Leigh}(1989)}]{Leigh:1989jq}%
  \BibitemOpen
  \bibfield  {author} {\bibinfo {author} {\bibfnamefont {R.~G.}\ \bibnamefont {Leigh}},\ }\bibfield  {title} {\enquote {\bibinfo {title} {{Dirac-Born-Infeld Action from Dirichlet Sigma Model}},}\ }\href {\doibase 10.1142/S0217732389003099} {\bibfield  {journal} {\bibinfo  {journal} {Mod. Phys. Lett. A}\ }\textbf {\bibinfo {volume} {4}},\ \bibinfo {pages} {2767} (\bibinfo {year} {1989})}\BibitemShut {NoStop}%
\bibitem [{\citenamefont {Ayon-Beato}\ and\ \citenamefont {Garcia}(1998)}]{Ayon-Beato:1998hmi}%
  \BibitemOpen
  \bibfield  {author} {\bibinfo {author} {\bibfnamefont {Eloy}\ \bibnamefont {Ayon-Beato}}\ and\ \bibinfo {author} {\bibfnamefont {Alberto}\ \bibnamefont {Garcia}},\ }\bibfield  {title} {\enquote {\bibinfo {title} {{Regular black hole in general relativity coupled to nonlinear electrodynamics}},}\ }\href {\doibase 10.1103/PhysRevLett.80.5056} {\bibfield  {journal} {\bibinfo  {journal} {Phys. Rev. Lett.}\ }\textbf {\bibinfo {volume} {80}},\ \bibinfo {pages} {5056--5059} (\bibinfo {year} {1998})},\ \Eprint {http://arxiv.org/abs/gr-qc/9911046} {arXiv:gr-qc/9911046} \BibitemShut {NoStop}%
\bibitem [{\citenamefont {Hale}\ \emph {et~al.}(2023)\citenamefont {Hale}, \citenamefont {Kubiz{\v{n}}{\'a}k}, \citenamefont {Sv{\'\i}tek},\ and\ \citenamefont {Tahamtan}}]{Hale:2023dpf}%
  \BibitemOpen
  \bibfield  {author} {\bibinfo {author} {\bibfnamefont {Tom{\'a}{\v{s}}}\ \bibnamefont {Hale}}, \bibinfo {author} {\bibfnamefont {David}\ \bibnamefont {Kubiz{\v{n}}{\'a}k}}, \bibinfo {author} {\bibfnamefont {Otakar}\ \bibnamefont {Sv{\'\i}tek}}, \ and\ \bibinfo {author} {\bibfnamefont {Tayebeh}\ \bibnamefont {Tahamtan}},\ }\bibfield  {title} {\enquote {\bibinfo {title} {{Solutions and basic properties of regularized Maxwell theory}},}\ }\href {\doibase 10.1103/PhysRevD.107.124031} {\bibfield  {journal} {\bibinfo  {journal} {Phys. Rev. D}\ }\textbf {\bibinfo {volume} {107}},\ \bibinfo {pages} {124031} (\bibinfo {year} {2023})},\ \Eprint {http://arxiv.org/abs/2303.16928} {arXiv:2303.16928 [gr-qc]} \BibitemShut {NoStop}%
\bibitem [{\citenamefont {Tahamtan}(2021)}]{Tahamtan:2020lvq}%
  \BibitemOpen
  \bibfield  {author} {\bibinfo {author} {\bibfnamefont {T.}~\bibnamefont {Tahamtan}},\ }\bibfield  {title} {\enquote {\bibinfo {title} {{Compatibility of nonlinear electrodynamics models with Robinson-Trautman geometry}},}\ }\href {\doibase 10.1103/PhysRevD.103.064052} {\bibfield  {journal} {\bibinfo  {journal} {Phys. Rev. D}\ }\textbf {\bibinfo {volume} {103}},\ \bibinfo {pages} {064052} (\bibinfo {year} {2021})},\ \Eprint {http://arxiv.org/abs/2010.01689} {arXiv:2010.01689 [gr-qc]} \BibitemShut {NoStop}%
\bibitem [{\citenamefont {Tahamtan}\ and\ \citenamefont {Sv{\'\i}tek}(2016)}]{Tahamtan:2015bha}%
  \BibitemOpen
  \bibfield  {author} {\bibinfo {author} {\bibfnamefont {T.}~\bibnamefont {Tahamtan}}\ and\ \bibinfo {author} {\bibfnamefont {O.}~\bibnamefont {Sv{\'\i}tek}},\ }\bibfield  {title} {\enquote {\bibinfo {title} {{Robinson{\textendash}Trautman solution with nonlinear electrodynamics}},}\ }\href {\doibase 10.1140/epjc/s10052-016-4175-9} {\bibfield  {journal} {\bibinfo  {journal} {Eur. Phys. J. C}\ }\textbf {\bibinfo {volume} {76}},\ \bibinfo {pages} {335} (\bibinfo {year} {2016})},\ \Eprint {http://arxiv.org/abs/1510.01183} {arXiv:1510.01183 [gr-qc]} \BibitemShut {NoStop}%
\bibitem [{\citenamefont {Kubiznak}\ \emph {et~al.}(2022)\citenamefont {Kubiznak}, \citenamefont {Tahamtan},\ and\ \citenamefont {Svitek}}]{Kubiznak:2022vft}%
  \BibitemOpen
  \bibfield  {author} {\bibinfo {author} {\bibfnamefont {David}\ \bibnamefont {Kubiznak}}, \bibinfo {author} {\bibfnamefont {Tayebeh}\ \bibnamefont {Tahamtan}}, \ and\ \bibinfo {author} {\bibfnamefont {Otakar}\ \bibnamefont {Svitek}},\ }\bibfield  {title} {\enquote {\bibinfo {title} {{Slowly rotating black holes in nonlinear electrodynamics}},}\ }\href {\doibase 10.1103/PhysRevD.105.104064} {\bibfield  {journal} {\bibinfo  {journal} {Phys. Rev. D}\ }\textbf {\bibinfo {volume} {105}},\ \bibinfo {pages} {104064} (\bibinfo {year} {2022})},\ \Eprint {http://arxiv.org/abs/2203.01919} {arXiv:2203.01919 [gr-qc]} \BibitemShut {NoStop}%
\bibitem [{\citenamefont {Kastor}\ \emph {et~al.}(2009)\citenamefont {Kastor}, \citenamefont {Ray},\ and\ \citenamefont {Traschen}}]{Kastor:2009wy}%
  \BibitemOpen
  \bibfield  {author} {\bibinfo {author} {\bibfnamefont {David}\ \bibnamefont {Kastor}}, \bibinfo {author} {\bibfnamefont {Sourya}\ \bibnamefont {Ray}}, \ and\ \bibinfo {author} {\bibfnamefont {Jennie}\ \bibnamefont {Traschen}},\ }\bibfield  {title} {\enquote {\bibinfo {title} {{Enthalpy and the Mechanics of AdS Black Holes}},}\ }\href {\doibase 10.1088/0264-9381/26/19/195011} {\bibfield  {journal} {\bibinfo  {journal} {Class. Quant. Grav.}\ }\textbf {\bibinfo {volume} {26}},\ \bibinfo {pages} {195011} (\bibinfo {year} {2009})},\ \Eprint {http://arxiv.org/abs/0904.2765} {arXiv:0904.2765 [hep-th]} \BibitemShut {NoStop}%
\bibitem [{\citenamefont {Toshmatov}\ \emph {et~al.}(2017)\citenamefont {Toshmatov}, \citenamefont {Bambi}, \citenamefont {Ahmedov}, \citenamefont {Abdujabbarov},\ and\ \citenamefont {Stuchl{\'\i}k}}]{Toshmatov:2017kmw}%
  \BibitemOpen
  \bibfield  {author} {\bibinfo {author} {\bibfnamefont {Bobir}\ \bibnamefont {Toshmatov}}, \bibinfo {author} {\bibfnamefont {Cosimo}\ \bibnamefont {Bambi}}, \bibinfo {author} {\bibfnamefont {Bobomurat}\ \bibnamefont {Ahmedov}}, \bibinfo {author} {\bibfnamefont {Ahmadjon}\ \bibnamefont {Abdujabbarov}}, \ and\ \bibinfo {author} {\bibfnamefont {Zden{\v{e}}k}\ \bibnamefont {Stuchl{\'\i}k}},\ }\bibfield  {title} {\enquote {\bibinfo {title} {{Energy conditions of non-singular black hole spacetimes in conformal gravity}},}\ }\href {\doibase 10.1140/epjc/s10052-017-5112-2} {\bibfield  {journal} {\bibinfo  {journal} {Eur. Phys. J. C}\ }\textbf {\bibinfo {volume} {77}},\ \bibinfo {pages} {542} (\bibinfo {year} {2017})},\ \Eprint {http://arxiv.org/abs/1702.06855} {arXiv:1702.06855 [gr-qc]} \BibitemShut {NoStop}%
\bibitem [{\citenamefont {Wei}\ \emph {et~al.}(2022{\natexlab{b}})\citenamefont {Wei}, \citenamefont {Liu},\ and\ \citenamefont {Mann}}]{29}%
  \BibitemOpen
  \bibfield  {author} {\bibinfo {author} {\bibfnamefont {Shao-Wen}\ \bibnamefont {Wei}}, \bibinfo {author} {\bibfnamefont {Yu-Xiao}\ \bibnamefont {Liu}}, \ and\ \bibinfo {author} {\bibfnamefont {Robert~B.}\ \bibnamefont {Mann}},\ }\bibfield  {title} {\enquote {\bibinfo {title} {{Black Hole Solutions as Topological Thermodynamic Defects}},}\ }\href {\doibase 10.1103/PhysRevLett.129.191101} {\bibfield  {journal} {\bibinfo  {journal} {Phys. Rev. Lett.}\ }\textbf {\bibinfo {volume} {129}},\ \bibinfo {pages} {191101} (\bibinfo {year} {2022}{\natexlab{b}})},\ \Eprint {http://arxiv.org/abs/2208.01932} {arXiv:2208.01932 [gr-qc]} \BibitemShut {NoStop}%
\bibitem [{\citenamefont {York}(1986)}]{york}%
  \BibitemOpen
  \bibfield  {author} {\bibinfo {author} {\bibfnamefont {James~W.}\ \bibnamefont {York}, \bibfnamefont {Jr.}},\ }\bibfield  {title} {\enquote {\bibinfo {title} {{Black hole thermodynamics and the Euclidean Einstein action}},}\ }\href {\doibase 10.1103/PhysRevD.33.2092} {\bibfield  {journal} {\bibinfo  {journal} {Phys. Rev. D}\ }\textbf {\bibinfo {volume} {33}},\ \bibinfo {pages} {2092--2099} (\bibinfo {year} {1986})}\BibitemShut {NoStop}%
\bibitem [{\citenamefont {Wei}\ and\ \citenamefont {Liu}(2022{\natexlab{b}})}]{Wei:2021vdx}%
  \BibitemOpen
  \bibfield  {author} {\bibinfo {author} {\bibfnamefont {Shao-Wen}\ \bibnamefont {Wei}}\ and\ \bibinfo {author} {\bibfnamefont {Yu-Xiao}\ \bibnamefont {Liu}},\ }\bibfield  {title} {\enquote {\bibinfo {title} {{Topology of black hole thermodynamics}},}\ }\href {\doibase 10.1103/PhysRevD.105.104003} {\bibfield  {journal} {\bibinfo  {journal} {Phys. Rev. D}\ }\textbf {\bibinfo {volume} {105}},\ \bibinfo {pages} {104003} (\bibinfo {year} {2022}{\natexlab{b}})},\ \Eprint {http://arxiv.org/abs/2112.01706} {arXiv:2112.01706 [gr-qc]} \BibitemShut {NoStop}%
\bibitem [{\citenamefont {Cai}\ and\ \citenamefont {Cho}(1999)}]{Cai:1998ep}%
  \BibitemOpen
  \bibfield  {author} {\bibinfo {author} {\bibfnamefont {R.~G.}\ \bibnamefont {Cai}}\ and\ \bibinfo {author} {\bibfnamefont {J.~H.}\ \bibnamefont {Cho}},\ }\bibfield  {title} {\enquote {\bibinfo {title} {Thermodynamic curvature of the btz black hole},}\ }\href {\doibase 10.1103/PhysRevD.60.067502} {\bibfield  {journal} {\bibinfo  {journal} {Phys. Rev. D}\ }\textbf {\bibinfo {volume} {60}},\ \bibinfo {pages} {067502} (\bibinfo {year} {1999})}\BibitemShut {NoStop}%
\bibitem [{\citenamefont {Wei}\ and\ \citenamefont {Liu}(2015)}]{Wei:2015iwa}%
  \BibitemOpen
  \bibfield  {author} {\bibinfo {author} {\bibfnamefont {S.~W.}\ \bibnamefont {Wei}}\ and\ \bibinfo {author} {\bibfnamefont {Y.~X.}\ \bibnamefont {Liu}},\ }\bibfield  {title} {\enquote {\bibinfo {title} {Insight into the microscopic structure of an ads black hole from a thermodynamical phase transition},}\ }\href {\doibase 10.1103/PhysRevLett.115.111302} {\bibfield  {journal} {\bibinfo  {journal} {Phys. Rev. Lett.}\ }\textbf {\bibinfo {volume} {115}},\ \bibinfo {pages} {111302} (\bibinfo {year} {2015})},\ \bibinfo {note} {erratum: Phys. Rev. Lett. 116, 169903 (2016), doi:10.1103/PhysRevLett.116.169903}\BibitemShut {NoStop}%
\bibitem [{\citenamefont {Wei}\ \emph {et~al.}(2019)\citenamefont {Wei}, \citenamefont {Liu},\ and\ \citenamefont {Mann}}]{Wei:2019uqg}%
  \BibitemOpen
  \bibfield  {author} {\bibinfo {author} {\bibfnamefont {S.~W.}\ \bibnamefont {Wei}}, \bibinfo {author} {\bibfnamefont {Y.~X.}\ \bibnamefont {Liu}}, \ and\ \bibinfo {author} {\bibfnamefont {R.~B.}\ \bibnamefont {Mann}},\ }\bibfield  {title} {\enquote {\bibinfo {title} {Repulsive interactions and universal properties of charged anti--de sitter black hole microstructures},}\ }\href {\doibase 10.1103/PhysRevLett.123.071103} {\bibfield  {journal} {\bibinfo  {journal} {Phys. Rev. Lett.}\ }\textbf {\bibinfo {volume} {123}},\ \bibinfo {pages} {071103} (\bibinfo {year} {2019})}\BibitemShut {NoStop}%
\bibitem [{\citenamefont {Guo}\ \emph {et~al.}(2019)\citenamefont {Guo}, \citenamefont {Li}, \citenamefont {Zhang},\ and\ \citenamefont {Zhao}}]{Guo:2019oad}%
  \BibitemOpen
  \bibfield  {author} {\bibinfo {author} {\bibfnamefont {X.~Y.}\ \bibnamefont {Guo}}, \bibinfo {author} {\bibfnamefont {H.~F.}\ \bibnamefont {Li}}, \bibinfo {author} {\bibfnamefont {L.~C.}\ \bibnamefont {Zhang}}, \ and\ \bibinfo {author} {\bibfnamefont {R.}~\bibnamefont {Zhao}},\ }\bibfield  {title} {\enquote {\bibinfo {title} {Microstructure and continuous phase transition of a reissner--nordström--ads black hole},}\ }\href {\doibase 10.1103/PhysRevD.100.064036} {\bibfield  {journal} {\bibinfo  {journal} {Phys. Rev. D}\ }\textbf {\bibinfo {volume} {100}},\ \bibinfo {pages} {064036} (\bibinfo {year} {2019})}\BibitemShut {NoStop}%
\bibitem [{\citenamefont {Xu}\ \emph {et~al.}(2020{\natexlab{a}})\citenamefont {Xu}, \citenamefont {Wu},\ and\ \citenamefont {Yang}}]{Xu:2020gud}%
  \BibitemOpen
  \bibfield  {author} {\bibinfo {author} {\bibfnamefont {Z.~M.}\ \bibnamefont {Xu}}, \bibinfo {author} {\bibfnamefont {B.}~\bibnamefont {Wu}}, \ and\ \bibinfo {author} {\bibfnamefont {W.~L.}\ \bibnamefont {Yang}},\ }\bibfield  {title} {\enquote {\bibinfo {title} {Ruppeiner thermodynamic geometry for the schwarzschild--ads black hole},}\ }\href {\doibase 10.1103/PhysRevD.101.024018} {\bibfield  {journal} {\bibinfo  {journal} {Phys. Rev. D}\ }\textbf {\bibinfo {volume} {101}},\ \bibinfo {pages} {024018} (\bibinfo {year} {2020}{\natexlab{a}})}\BibitemShut {NoStop}%
\bibitem [{\citenamefont {Ghosh}\ and\ \citenamefont {Bhamidipati}(2020{\natexlab{b}})}]{Ghosh:2020kba}%
  \BibitemOpen
  \bibfield  {author} {\bibinfo {author} {\bibfnamefont {A.}~\bibnamefont {Ghosh}}\ and\ \bibinfo {author} {\bibfnamefont {C.}~\bibnamefont {Bhamidipati}},\ }\bibfield  {title} {\enquote {\bibinfo {title} {Thermodynamic geometry and interacting microstructures of btz black holes},}\ }\href {\doibase 10.1103/PhysRevD.101.106007} {\bibfield  {journal} {\bibinfo  {journal} {Phys. Rev. D}\ }\textbf {\bibinfo {volume} {101}},\ \bibinfo {pages} {106007} (\bibinfo {year} {2020}{\natexlab{b}})}\BibitemShut {NoStop}%
\bibitem [{\citenamefont {Xu}\ \emph {et~al.}(2020{\natexlab{b}})\citenamefont {Xu}, \citenamefont {Wu},\ and\ \citenamefont {Yang}}]{Xu:2020ftx}%
  \BibitemOpen
  \bibfield  {author} {\bibinfo {author} {\bibfnamefont {Z.~M.}\ \bibnamefont {Xu}}, \bibinfo {author} {\bibfnamefont {B.}~\bibnamefont {Wu}}, \ and\ \bibinfo {author} {\bibfnamefont {W.~L.}\ \bibnamefont {Yang}},\ }\bibfield  {title} {\enquote {\bibinfo {title} {Diagnosis inspired by the thermodynamic geometry for different thermodynamic schemes of the charged btz black hole},}\ }\href {\doibase 10.1140/epjc/s10052-020-08501-y} {\bibfield  {journal} {\bibinfo  {journal} {Eur. Phys. J. C}\ }\textbf {\bibinfo {volume} {80}},\ \bibinfo {pages} {997} (\bibinfo {year} {2020}{\natexlab{b}})}\BibitemShut {NoStop}%
\bibitem [{\citenamefont {Hirunsirisawat}\ \emph {et~al.}(2022)\citenamefont {Hirunsirisawat}, \citenamefont {Nakarachinda},\ and\ \citenamefont {Promsiri}}]{Prom}%
  \BibitemOpen
  \bibfield  {author} {\bibinfo {author} {\bibfnamefont {E.}~\bibnamefont {Hirunsirisawat}}, \bibinfo {author} {\bibfnamefont {R.}~\bibnamefont {Nakarachinda}}, \ and\ \bibinfo {author} {\bibfnamefont {C.}~\bibnamefont {Promsiri}},\ }\bibfield  {title} {\enquote {\bibinfo {title} {Emergent phase, thermodynamic geometry, and criticality of charged black holes from r\'enyi statistics},}\ }\href {\doibase 10.1103/PhysRevD.105.124049} {\bibfield  {journal} {\bibinfo  {journal} {Phys. Rev. D}\ }\textbf {\bibinfo {volume} {105}},\ \bibinfo {pages} {124049} (\bibinfo {year} {2022})}\BibitemShut {NoStop}%
\bibitem [{\citenamefont {Dehghani}\ \emph {et~al.}(2023)\citenamefont {Dehghani}, \citenamefont {Pourhassan}, \citenamefont {Zarepour},\ and\ \citenamefont {Saridakis}}]{Dehghani:2023yph}%
  \BibitemOpen
  \bibfield  {author} {\bibinfo {author} {\bibfnamefont {A.}~\bibnamefont {Dehghani}}, \bibinfo {author} {\bibfnamefont {B.}~\bibnamefont {Pourhassan}}, \bibinfo {author} {\bibfnamefont {S.}~\bibnamefont {Zarepour}}, \ and\ \bibinfo {author} {\bibfnamefont {E.~N.}\ \bibnamefont {Saridakis}},\ }\bibfield  {title} {\enquote {\bibinfo {title} {Thermodynamic schemes of charged btz-like black holes in arbitrary dimensions},}\ }\href {\doibase 10.1103/PhysRevD.108.024006} {\bibfield  {journal} {\bibinfo  {journal} {Phys. Rev. D}\ }\textbf {\bibinfo {volume} {108}},\ \bibinfo {pages} {024006} (\bibinfo {year} {2023})}\BibitemShut {NoStop}%
\bibitem [{\citenamefont {{Suleimanov}}\ \emph {et~al.}(2025)\citenamefont {{Suleimanov}}, \citenamefont {{Nosirov}}, \citenamefont {{Yusupov}}, \citenamefont {{Chaves}}, \citenamefont {{Berdiyorov}},\ and\ \citenamefont {{Rakhimov}}}]{2025PhyB..71417484S}%
  \BibitemOpen
  \bibfield  {author} {\bibinfo {author} {\bibfnamefont {M.~M.}\ \bibnamefont {{Suleimanov}}}, \bibinfo {author} {\bibfnamefont {M.~U.}\ \bibnamefont {{Nosirov}}}, \bibinfo {author} {\bibfnamefont {H.~T.}\ \bibnamefont {{Yusupov}}}, \bibinfo {author} {\bibfnamefont {A.}~\bibnamefont {{Chaves}}}, \bibinfo {author} {\bibfnamefont {G.~R.}\ \bibnamefont {{Berdiyorov}}}, \ and\ \bibinfo {author} {\bibfnamefont {Kh.~Yu.}\ \bibnamefont {{Rakhimov}}},\ }\bibfield  {title} {\enquote {\bibinfo {title} {{Wave-packet dynamics in monolayer graphene with periodic scattering potentials}},}\ }\href {\doibase 10.1016/j.physb.2025.417484} {\bibfield  {journal} {\bibinfo  {journal} {Physica B Condensed Matter}\ }\textbf {\bibinfo {volume} {714}},\ \bibinfo {eid} {417484} (\bibinfo {year} {2025})},\ \Eprint {http://arxiv.org/abs/2411.02896} {arXiv:2411.02896 [cond-mat.mes-hall]} \BibitemShut {NoStop}%
\bibitem [{\citenamefont {Ghaffari}\ \emph {et~al.}(2024)\citenamefont {Ghaffari}, \citenamefont {Luciano},\ and\ \citenamefont {Sheykhi}}]{Ghaffari:2023vcw}%
  \BibitemOpen
  \bibfield  {author} {\bibinfo {author} {\bibfnamefont {S.}~\bibnamefont {Ghaffari}}, \bibinfo {author} {\bibfnamefont {G.~G.}\ \bibnamefont {Luciano}}, \ and\ \bibinfo {author} {\bibfnamefont {A.}~\bibnamefont {Sheykhi}},\ }\bibfield  {title} {\enquote {\bibinfo {title} {Nonextensive entropies impact onto thermodynamics and phase structure of kerr--newman black holes},}\ }\href {\doibase 10.1016/j.dark.2024.101447} {\bibfield  {journal} {\bibinfo  {journal} {Phys. Dark Univ.}\ }\textbf {\bibinfo {volume} {44}},\ \bibinfo {pages} {101447} (\bibinfo {year} {2024})}\BibitemShut {NoStop}%
\bibitem [{\citenamefont {Sekhmani}\ \emph {et~al.}(2024)\citenamefont {Sekhmani}, \citenamefont {Luciano}, \citenamefont {Rayimbaev}, \citenamefont {Jasim}, \citenamefont {Al-Badawi},\ and\ \citenamefont {Maurya}}]{Sekhmani:2024udl}%
  \BibitemOpen
  \bibfield  {author} {\bibinfo {author} {\bibfnamefont {Y.}~\bibnamefont {Sekhmani}}, \bibinfo {author} {\bibfnamefont {G.~G.}\ \bibnamefont {Luciano}}, \bibinfo {author} {\bibfnamefont {J.}~\bibnamefont {Rayimbaev}}, \bibinfo {author} {\bibfnamefont {M.~K.}\ \bibnamefont {Jasim}}, \bibinfo {author} {\bibfnamefont {A.}~\bibnamefont {Al-Badawi}}, \ and\ \bibinfo {author} {\bibfnamefont {S.~K.}\ \bibnamefont {Maurya}},\ }\bibfield  {title} {\enquote {\bibinfo {title} {Topological ads black holes surrounded by chaplygin dark fluid: From stability to geometrothermodynamic analysis},}\ }\href {\doibase 10.1016/j.dark.2024.101567} {\bibfield  {journal} {\bibinfo  {journal} {Phys. Dark Univ.}\ }\textbf {\bibinfo {volume} {46}},\ \bibinfo {pages} {101567} (\bibinfo {year} {2024})}\BibitemShut {NoStop}%
\bibitem [{\citenamefont {Quevedo}(2007)}]{Quevedo:2006xk}%
  \BibitemOpen
  \bibfield  {author} {\bibinfo {author} {\bibfnamefont {H.}~\bibnamefont {Quevedo}},\ }\bibfield  {title} {\enquote {\bibinfo {title} {{Geometrothermodynamics}},}\ }\href {\doibase 10.1063/1.2409524} {\bibfield  {journal} {\bibinfo  {journal} {J. Math. Phys.}\ }\textbf {\bibinfo {volume} {48}},\ \bibinfo {pages} {013506} (\bibinfo {year} {2007})}\BibitemShut {NoStop}%
\bibitem [{\citenamefont {Quevedo}(2008)}]{Quevedo:2007mj}%
  \BibitemOpen
  \bibfield  {author} {\bibinfo {author} {\bibfnamefont {H.}~\bibnamefont {Quevedo}},\ }\bibfield  {title} {\enquote {\bibinfo {title} {{Geometrothermodynamics of black holes}},}\ }\href {\doibase 10.1007/s10714-007-0586-0} {\bibfield  {journal} {\bibinfo  {journal} {Gen. Rel. Grav.}\ }\textbf {\bibinfo {volume} {40}},\ \bibinfo {pages} {971--984} (\bibinfo {year} {2008})},\ \Eprint {http://arxiv.org/abs/0704.3102} {arXiv:0704.3102 [gr-qc]} \BibitemShut {NoStop}%
\bibitem [{\citenamefont {Chaudhary}\ \emph {et~al.}(2025)\citenamefont {Chaudhary}, \citenamefont {Anwar}, \citenamefont {Atamurotov}, \citenamefont {Mubaraki},\ and\ \citenamefont {Alam}}]{qq1}%
  \BibitemOpen
  \bibfield  {author} {\bibinfo {author} {\bibfnamefont {Shahid}\ \bibnamefont {Chaudhary}}, \bibinfo {author} {\bibfnamefont {Talha}\ \bibnamefont {Anwar}}, \bibinfo {author} {\bibfnamefont {Farruh}\ \bibnamefont {Atamurotov}}, \bibinfo {author} {\bibfnamefont {Ali~M.}\ \bibnamefont {Mubaraki}}, \ and\ \bibinfo {author} {\bibfnamefont {M.~M.}\ \bibnamefont {Alam}},\ }\bibfield  {title} {\enquote {\bibinfo {title} {{Gravitational lensing and shadows of dilatonic black holes in dilaton-massive gravity}},}\ }\href {\doibase 10.1016/j.nuclphysb.2025.117075} {\bibfield  {journal} {\bibinfo  {journal} {Nucl. Phys. B}\ }\textbf {\bibinfo {volume} {1018}},\ \bibinfo {pages} {117075} (\bibinfo {year} {2025})}\BibitemShut {NoStop}%
\bibitem [{\citenamefont {Mustafa}\ \emph {et~al.}(2025)\citenamefont {Mustafa}, \citenamefont {Alimova}, \citenamefont {Atamurotov}, \citenamefont {Ibraheem}, \citenamefont {Channuie},\ and\ \citenamefont {Bahaddinova}}]{qq2}%
  \BibitemOpen
  \bibfield  {author} {\bibinfo {author} {\bibfnamefont {G.}~\bibnamefont {Mustafa}}, \bibinfo {author} {\bibfnamefont {Asalkhon}\ \bibnamefont {Alimova}}, \bibinfo {author} {\bibfnamefont {Farruh}\ \bibnamefont {Atamurotov}}, \bibinfo {author} {\bibfnamefont {Awad~A.}\ \bibnamefont {Ibraheem}}, \bibinfo {author} {\bibfnamefont {Phongpichit}\ \bibnamefont {Channuie}}, \ and\ \bibinfo {author} {\bibfnamefont {Gunel}\ \bibnamefont {Bahaddinova}},\ }\bibfield  {title} {\enquote {\bibinfo {title} {{Testing regular black holes in the framework of asymptotically safe gravity using particle dynamics, QPOs, and shadow constraints}},}\ }\href {\doibase 10.1140/epjc/s10052-025-14431-3} {\bibfield  {journal} {\bibinfo  {journal} {Eur. Phys. J. C}\ }\textbf {\bibinfo {volume} {85}},\ \bibinfo {pages} {741} (\bibinfo {year} {2025})}\BibitemShut {NoStop}%
\bibitem [{\citenamefont {Turimov}\ \emph {et~al.}(2025{\natexlab{a}})\citenamefont {Turimov}, \citenamefont {Usanov},\ and\ \citenamefont {Khamroev}}]{qq3}%
  \BibitemOpen
  \bibfield  {author} {\bibinfo {author} {\bibfnamefont {Bobur}\ \bibnamefont {Turimov}}, \bibinfo {author} {\bibfnamefont {Sulton}\ \bibnamefont {Usanov}}, \ and\ \bibinfo {author} {\bibfnamefont {Yokubjon}\ \bibnamefont {Khamroev}},\ }\bibfield  {title} {\enquote {\bibinfo {title} {{Particles acceleration by Bocharova{\textendash}Bronnikov{\textendash}Melnikov{\textendash}Bekenstein black hole}},}\ }\href {\doibase 10.1016/j.dark.2025.101876} {\bibfield  {journal} {\bibinfo  {journal} {Phys. Dark Univ.}\ }\textbf {\bibinfo {volume} {48}},\ \bibinfo {pages} {101876} (\bibinfo {year} {2025}{\natexlab{a}})},\ \Eprint {http://arxiv.org/abs/2502.11185} {arXiv:2502.11185 [gr-qc]} \BibitemShut {NoStop}%
\bibitem [{\citenamefont {Turimov}\ \emph {et~al.}(2025{\natexlab{b}})\citenamefont {Turimov}, \citenamefont {Urinov}, \citenamefont {Mardiev}, \citenamefont {Usanov}, \citenamefont {Akylbayev},\ and\ \citenamefont {Kenzhebekova}}]{Turimov:2025qkz}%
  \BibitemOpen
  \bibfield  {author} {\bibinfo {author} {\bibfnamefont {Bobur}\ \bibnamefont {Turimov}}, \bibinfo {author} {\bibfnamefont {Sunnatillo}\ \bibnamefont {Urinov}}, \bibinfo {author} {\bibfnamefont {Islomkhon}\ \bibnamefont {Mardiev}}, \bibinfo {author} {\bibfnamefont {Sulton}\ \bibnamefont {Usanov}}, \bibinfo {author} {\bibfnamefont {Musabek}\ \bibnamefont {Akylbayev}}, \ and\ \bibinfo {author} {\bibfnamefont {Rabiga}\ \bibnamefont {Kenzhebekova}},\ }\bibfield  {title} {\enquote {\bibinfo {title} {{Tidal forces in Bocharova{\textendash}Bronnikov{\textendash}Melnikov{\textendash}Bekenstein spacetime}},}\ }\href {\doibase 10.1007/s40065-025-00573-5} {\  (\bibinfo {year} {2025}{\natexlab{b}}),\ 10.1007/s40065-025-00573-5}\BibitemShut {NoStop}%
\bibitem [{\citenamefont {Yunusov}\ \emph {et~al.}(2025)\citenamefont {Yunusov}, \citenamefont {Turimov}, \citenamefont {Khamroev}, \citenamefont {Usanov}, \citenamefont {Turaev},\ and\ \citenamefont {Kuliyeva}}]{Yunusov:2025chw}%
  \BibitemOpen
  \bibfield  {author} {\bibinfo {author} {\bibfnamefont {Odil}\ \bibnamefont {Yunusov}}, \bibinfo {author} {\bibfnamefont {Bobur}\ \bibnamefont {Turimov}}, \bibinfo {author} {\bibfnamefont {Yokubjon}\ \bibnamefont {Khamroev}}, \bibinfo {author} {\bibfnamefont {Sulton}\ \bibnamefont {Usanov}}, \bibinfo {author} {\bibfnamefont {Farkhodjon}\ \bibnamefont {Turaev}}, \ and\ \bibinfo {author} {\bibfnamefont {Markhabo}\ \bibnamefont {Kuliyeva}},\ }\bibfield  {title} {\enquote {\bibinfo {title} {{Charged Zipoy{\textendash}Voorhees metric in string theory}},}\ }\href {\doibase 10.1016/j.aop.2025.170151} {\bibfield  {journal} {\bibinfo  {journal} {Annals Phys.}\ }\textbf {\bibinfo {volume} {481}},\ \bibinfo {pages} {170151} (\bibinfo {year} {2025})}\BibitemShut {NoStop}%
\end{thebibliography}%

\end{document}